\documentclass[preprint,aps,nopacs,preprintnumbers,amsmath,amssymb]{revtex4}

\usepackage{amsmath}

\usepackage{graphicx}
\usepackage{dcolumn}
\usepackage{bm}
\usepackage{color}

\usepackage{amsmath}
\usepackage{amssymb}

\def\be{\begin{equation}}
\def\en{\end{equation}}
\def\bea{\begin{eqnarray}}
\def\ena{\end{eqnarray}}
\def\bec{\begin{equation}\begin{array}{rcl}}
\def\p{\partial}

\def\gs{\gtrsim}
\def\ls{\lesssim}

\newcommand{\av}[1]{\langle{#1}\rangle}

\newcommand{\bi}[1]{\mbox{\boldmath$#1$}}

\newcommand{\ppp}[3]{{\bigg(}\frac{\partial {#1}}{\partial {#2}}{\bigg )}_{#3}}

\begin{document}
\title{Extension of  Kirkwood-Buff theory:  
Partial enthalpies, fluctuations of   energy density, 
 temperature, and pressure,  and solute-induced  effects in a mixture solvent }
\author{Akira Onuki\footnote{e-mail: onuki@scphys.kyoto-u.ac.jp 
(corresponding author)}}

\affiliation{
 Department of Physics, Kyoto University, Kyoto 606-8502, Japan 
}


\date{\today}

\begin{abstract} 
We present a statistical mechanical  theory of  multi-component fluids, 
where we consider   the correlation functions  of 
 the number  densities and the energy density in the grand canonical 
ensemble. In terms of their  space  integrals 
we express 
 the partial  volumes ${\bar v}_i$, the partial  enthalpies ${\bar H}_i$,  
 and other thermodynamic derivatives. These   ${\bar v}_i$ and  ${\bar H}_i$   
assume simple forms for binary mixtures 
and for ternary mixtures with a dilute solute. 
They are then related to  the space-dependent thermal fluctuations 
of the temperature and the pressure. The space averages of 
 these fluctuations are those  
introduced   by    Landau and Lifshits in the isothermal-isobaric ($T$-$p$) 
 ensemble. We also give expressions 
for the long-range  (nonlocal) correlations 
in the canonical and $T$-$p$ ensembles, which are 
inversely proportional to the system volume. 
For   a  mixture solvent, 
we  examine    
the solvent-induced solute-solute 
attraction and  the osmotic  enthalpy changes   
due to the solute doping   
using the  correlation function integrals. 
\end{abstract}


\maketitle


\noindent{\bf 1. Introduction}\\ 
The Kirkwood-Buff (KB) 
integrals $G_{ij}$ \cite{Buff,Naim2,Naim3,Naim,Oc} 
have been  used extensively 
in the investigation of   the molecular interactions  
in multi-component  fluids.  
For  species $i$ and $j$, $G_{ij}$ are  
  the space integrals 
of $g_{ij}(r)-1$,  where $g_{ij}(r)$ are the radial distribution functions. 
They have been determined experimentally 
by  the  small-angle   scattering  of $x$-ray 
and  neutrons  from  binary mixtures 
\cite{Nishi,Misawa,YKoga,Lep}, 
where   the combination $G_{11}+ G_{22}-2G_{12}$ represents 
asymmetry of the molecular interactions and its large size indicates 
enhancement of the concentration fluctuation \cite{Bagchi1,Ruck,Koga}. 
Ghosh and Bagchi \cite{Bagchi1}  numerically studied such 
clustering dynamically. 
In biology \cite{Pa,Tim,Smith,Smith1,Record,Shimizu,Mochi}, 
attention has been paid to the action of 
a cosolvent  on the  solubility and stability  
of proteins  in a primary solvent. 
In such solutions,  the solute interacts with 
the two solvent species differently 
 and the difference 
$G_{32}-G_{31}$ represents the strength of 
preferential solvation, 
where the index 3 refers to the solute. 
This difference grows with increasing 
 the  solute size  above  the solvent ones.

The partial volumes  ${\bar v}_i$ 
 have been measured (including those of  ions in water)
 \cite{Jones,Naka,Craig}. They  are  characteristic  volumes per particle  
satisfying  the sum rule ${\sum}_i {\bar v}_i n_i=1$, where  
  $n_i$ are  the average number densities. 
They can be expressed in terms of $G_{ij}$ \cite{Buff,Naim2,Naim,Naim3,Oc}. 
 On the other hand, the partial enthalpies   ${\bar H}_i$ 
are theoretically difficult quantities beyond  the 
Kirkwood-Buff theory.  They are defined as  
the  enthalpy change  upon 
 addition of a single particle of the $i$-th species 
at fixed temperature $T$ and pressure $p$ in a macroscopic solution. 
Thus,  ${\bar H}_i$  were  measured by calorimetric methods 
\cite{Hepler,Nil,YKoga1,Maham}. In chemical reactions, 
the reaction enthalpy is a linear combination of 
 the partial enthalpies 
of the reactant and product particles \cite{Landau-s,Callen}.
As numerical calculations of  ${\bar H}_i$,  
Frenkel's group  \cite{Frenkel} 
 used  a particle-insertion technique, while  
Vlugt's group \cite{Bedeaux,Vlugt,Vlugt1,VlugtF}  
  used   some  thermodynamic relations  
of the energy density. 
In this paper, we consider the correlation functions 
$  \av{\delta{\hat e}({\bi r})\delta {\hat n}_i({\bi 0})}$  
of   the  fluctuation of the energy density 
$\delta {\hat e}({\bi r})$ 
and those of the number densities 
$\delta {\hat n}_i({\bi r})$ 
in the grand-canonical ensemble. 
We write their space integrals  as $n_iJ_i$ and   
  express the partial enthalpies 
 in terms of $J_i$ and $G_{ij}$.

In this paper, we  introduce   space-dependent  
  temperature and  pressure fluctuations \cite{Landau-s,Onukibook}, 
 written as $\delta {\hat T}({\bi r})$ and $\delta {\hat p}({\bi r})$, 
respectively. They  are linear combinations 
of $\delta {\hat e}({\bi r})$ 
and   $\delta {\hat n}_i({\bi r})$,  
 where the coefficients in front of $\delta {\hat n}_i({\bi r})$ 
are thermodynamic derivatives closely   related to ${\bar H}_i$. 
We also introduce the fluctuation of the volume fraction 
 $\delta {\hat \phi}({\bi r})= {\sum}_i {\bar v}_i
\delta {\hat n}_i({\bi r})$ using ${\bar v}_i$ \cite{Kirk3,Oka1},  
 which can  be expressed as a linear combination  of 
  $\delta {\hat T}({\bi r})$ and $\delta {\hat p}({\bi r})$. 
We then calculate  the correlation function integrals 
among    $\delta {\hat n}_i({\bi r})$, $\delta {\hat e}({\bi r})$, 
$\delta {\hat T}({\bi r})$,  $\delta {\hat p}({\bi r})$, 
$\delta {\hat \phi}({\bi r})$, 
and the microscopically defined local 
stress tensor ${\hat{\Pi}}_{\alpha\beta}({\bi r})$ ($\alpha,\beta=x,y,z$). 
Landau and Lifshitz  \cite{Landau-s} introduced 
  temperature and  pressure fluctuations, $\Delta T$ and $\Delta p$,  
 in a finite system in the isothermal-isobaric ($T$-$p$) 
ensemble \cite{Kubo,Binderbook,Fre}, 
which are the space averages of 
 $\delta {\hat T}({\bi r})$ and $\delta {\hat p}({\bi r})$ 
in the container. 
We also give expressions for the   long-range (nonlocal) correlations 
in the  $T$-$p$ ensemble, which are  
well known for  the canonical ensemble \cite{Lebo}.

For binary systems, 
${\bar v}_i$ are known to have  simple expressions  in terms of 
 $G_{jk}$ \cite{Buff,Naim,Naim2,Oc}. 
In this paper, we  express ${\bar H}_i$ 
in simple forms in terms of $J_j$ and $G_{jk}$. 
For ternary solutions composed of  a mixture   solvent 
 and a dilute solute,  
we present  expressions of    ${\bar v}_3$ and  ${\bar H}_3$ 
for the solute  in terms of  
 $J_i$ and $G_{ij}$ up to the linear order in 
the solute density $n_3$. These expressions 
 are further  simplified in the 
infinite-dilution  limit $n_3\to 0$. The critical behaviors of 
 ${\bar v}_i$ and  ${\bar H}_i$  are  known unambiguously 
from their correlation function expressions.

For   ternary solutions, 
we    present a   theory on 
the  solvent-mediated,  solute-solute attraction 
 by calculating $G_{33}$ for the solute. It 
   arises from  the solute-induced  
solvent density changes, which are divided into those of 
$\delta{\hat \phi}({\bi r})$ \cite{Chan,Ben,Lev,Oo,Jiang,Oka1,Bagchi} 
 and those of the concentration deviation $\delta{\hat X}({\bi r})= 
  [n_1\delta{\hat n}_2({\bi r})-n_2 \delta{\hat n}_1({\bi r})]/n^2$ 
 \cite{Hori,Zemb1,Oka1}. As a related problem, 
 Asakura and  Oosawa presented  an   attractive 
potential between colloidal particles 
due to polymer depletion in  solvent-polymer-colloid mixtures 
 \cite{Oosawa,Vrij,Binder,Lek1}.   
 Finally, we examine the 
thermodynamics of a ternary solution  
in  osmotic equilibrium  with a reservoir at fixed 
 solvent chemical potentials $\mu_1$ and $\mu_2$. In this situation,   
   the osmotic pressure $\Pi$ 
has been investigated \cite{Mc,Koga,Ruck}.   
  We further calculate the 
enthalpy change $\Delta H$ 
as a function of  the solute density $n_3$.

 The organization of this paper is as follows. In Section 2, 
 we will present  thermodynamic relations of  the partial 
quantities. In Section 3, we will introduce 
the correlation function integrals.  
In Section 4, we will discuss the fluctuations 
in the temperature and the pressure. 
In Section 5, the partial quantities in binary mixtures will 
be calculated. In Section 6, the partial quantities 
and the solvent-mediated interaction 
will be discussed in ternary mixtures.  In section 7, the thermodynamic 
of ternary mixtures  in  osmotic 
equilibrium will be investigated.  

\vspace{2mm}
\noindent{\bf  2. Thermodynamics of multi-component fluids}\\
\vspace{2mm}
\noindent{\bf 2.1.  Partial quantities at fixed $p$ and $T$  }\\ 
We consider an equilibrium   fluid mixture  composed of    
$m$ components  without electrostatic interactions 
in a macroscopic  volume $V$. 
We assume no  applied field such as gravity and electric field 
and neglect the boundary effects. 
In this section, we present  fundamental thermodynamic relations 
needed in Sections 3-7. 

Let us consider   an  extensive 
 quantity $Q$  as a function of the temperature $T$, 
the pressure $p$, and the particle numbers $N_i=Vn_i$ ($1\le i\le m)$ 
with $n_i$ being the average densities.   
Its  partial quantities per particle are defined by   
\be 
{\bar Q}_i= \ppp{ Q}{ N_i}{T,p, N_{j\neq i}}, 
\en 
where we  fix  $T$, 
 $p$,  and $N_j$ with $j\neq i$ in the derivative. 
The extensivity of $Q$ means the relation 
$Q(T, p, \{\lambda N\})= 
\lambda Q(T,p,\{ N\})$ for $N_i\lambda\gg 1$, which  
yields the sum rule,   
\be 
Q= {\sum}_i {\bar Q}_i N_i=V{\sum}_i {\bar Q}_i n_i.
\en  
For the volume $V$ and the total entropy $S$,  
 the partial volumes ${\bar v}_i$ and 
the partial entropies ${\bar S}_i$ are written as  
\be 
{\bar v}_i= \ppp{ V}{ N_i}{T, p, N_{j\neq i}},~~
{\bar S}_i= \ppp{ S}{ N_i}{T, p,N_{j\neq i}}. 
\en 
 For the  Gibbs  free energy $G$, its partial 
quantities  are   equal to the chemical potentials $\mu_i$ 
from its  differential equation, 
\be 
dG= {\sum}_i \mu_i dN_i-SdT+ Vdp. 
\en  
Therefore, from  the enthalpy relation $H= G+TS$,  
the   partial enthalpies     are given by 
\be 
{\bar H}_i= \mu_i+ T{\bar S}_i.
\en

Varying $(\{N\},T,p)$  we  have the differential equation, 
\be
dQ=  {\sum}_i {\bar Q}_i dN_i+\ppp{Q}{T}{p, \{N\}} \hspace{-1mm}dT
 +\ppp{Q}{p}{T, \{N\}} \hspace{-1mm}
dp .
\en   
For $Q=V$ and $S$ the above relation is written as  
\bea 
&&\hspace{-10mm}
dV=  {\sum}_i {\bar v}_i dN_i  + V\alpha_pdT -V \kappa_T dp,\\  
&&\hspace{-10mm}
dS=  {\sum}_i {\bar S}_i dN_i+ VC_pdT/T -V \alpha_p dp , 
\ena 
where     $\alpha_p$ is the isobaric thermal expansivity, 
 $\kappa_T$ is the isothermal compressibility, and 
$ C_p$ is the isobaric heat capacity per unit volume. In the 
thermodynamic limit $V\to\infty$, we have  
\bea
&&\hspace{-15mm}
\alpha_p=-\frac{1}{n}\ppp{ n}{ T}{p,\{ x\}},~~~
\kappa_T=\frac{1}{n}\ppp{ n}{ p}{T,\{ x\}},~~~
{C_p}={ nT}\ppp{ \sigma}{ T}{p,\{ x\}} ,
\ena 
where  $n=\sum_i n_i=N/V$ and   $N={\sum}_iN_i$. 
Here,  $\sigma=S/N$ is  the entropy per particle 
and is related to  the entropy density by 
$s=S/V=n\sigma$, so $VC_p= T (\p S/\p T)_{p,\{N\}}
= NT  (\p \sigma/\p T)_{p,\{x\}}$. 
Note that the thermodynamic derivatives of 
intensive quantities   at fixed $ \{ N\}$ are 
those at fixed molar concentrations  $ \{ x\}$.  
In this paper,  $\{ x\}$ denotes the set of   
 $x_i= N_i/N= n_i/n$, 
while $\{ n\}$ denotes  the set of $n_i$.

We further consider  the density variable $q\equiv Q/V$.  
From Eqs.(6) and (7),  it obeys      
\be
dq= { \sum}_i {\bar Q}_i dn_i  +\ppp{q}{T}{p, \{x\}}\hspace{-1mm} dT 
+ \ppp{q}{p}{T, \{x\}}\hspace{-1mm}
 dp + q(\alpha_p dT-\kappa_T dp) .
\en  
 Since $q=1$ for $Q=V$, Eq.(10) yields the differential equation 
\cite{Oka1},         
\be 
   {\sum}_i {\bar v}_i dn_i +  \alpha_p  dT-\kappa_T dp=0,   
\en  
where the first term is the incremental 
change in  the particle 
volume fraction (see Eqs.(38) and (47)). 
Then,  we have two expressions for  ${\bar v}_i$:   
\be 
{\bar v}_i= \kappa_T  \ppp{p}{n_i}{T, n_{j\neq i} },~~
{\bar v}_i= -\alpha_p  \ppp{T}{n_i}{p, n_{j\neq i} }.   
\en 
The first relation  is convenient 
for dilute mixtures, since   $p$ 
can be   expanded with respect to 
the solute densities (see Section 6). 
The second relation follows from the first 
since  $ \alpha_p= \kappa_T (\p p/\p T)_{\{n\}}$ 
(see Eq.(27)).

Next, we  define  a symmetric matrix  
  $\{I_{ij}\}$  and its inverse matrix   $\{I^{ij}\}$  as 
\be 
\hspace{-0.112mm} 
I_{ij}= {k_BT} 
\ppp{ n_j}{\mu_i}{T, \mu_{k\neq i}}\hspace{-1mm}, ~
I^{ij}= \frac{1}{k_BT} 
\ppp{ \mu_j}{ n_i}{T, n_{k\neq i}} \hspace{-1mm},   
\en 
where $T$ is fixed and ${\sum}_k I_{ik}I^{kj} =\delta_{ij}$.      
Using the Gibbs-Duhem relation 
 in the first relation in Eq.(12), 
we can relate $\{ {\bar v}\}$ and $\{ {n}\}$ as 
\be 
{\bar v}_i= k_BT \kappa_T  {\sum}_{j} I^{ij} n_j, ~~
 {\sum}_j I_{ij} {\bar v}_j =k_BT\kappa_Tn_i .
\en 
The isothermal compressibility  $\kappa_T$ is then expressed as  
\be 
k_BT {\kappa_T}= 1{\big /}\big[ {\sum}_{i,j} n_i n_j I^{ij}\big]
= {\sum}_{i,j} {\bar v}_i 
{\bar v}_j I_{ij} .  
\en

We note that  Eq.(4) yields  the well-known expression,  
\be 
{\bar v}_i= \ppp{ \mu_i}{ p}{T, \{ x\}}
=k_BT{\sum}_j I^{ij} \ppp{ n_j}{p}{T, \{ x\}} . 
\en 
which is equivalent to	   Eq.(12) 
from $(\p n_j/\p p)_{T,\{x\}}= n_j \kappa_T$ for any $j$. 
Then,  Eq.(4)  also gives   
\be 
{\bar S}_i= - \ppp{\mu_i}{T}{p,\{x\}},~~
 {\bar H}_i= -T^2 \ppp{(\mu_i/T)}{T}{p,\{x\}}. 
\en 
Here,  we have $(\p w/\p T)_{p,\{x\}}
= (\p w/\p T)_{\{n \}}-\alpha_p 
n(\p w/\p n)_{T,\{x\}}$ for any $w$.  

\vspace{2mm}
\noindent{\bf 2.2.  Derivatives of energy density with 
respect to $n_i$ }\\ 
We next introduce the thermodynamic 
 derivatives of  the internal  energy density $e$ 
with respect to the number densities $n_i$:  
\bea 
&&\hspace{-15mm}
M_i =\ppp{e}{ n_i}{T,n_{j\neq i}} =
- T^2\ppp{ (\mu_i/T) }{ T}{\{ n\}},\\
&&\hspace{-15mm}
R_i =\ppp{e}{ n_i}{p,n_{j\neq i}} =M_i 
- \frac{1}{\alpha_p} C_V {\bar v}_i , 
\ena  
where we fix $T $  in   $M_i$ 
and $p$ in $R_i$. In Eq.(18) use is made of    
$(\p s/\p n_i)_{T,n_{j\neq i}}=- 
(\p \mu_i/\p T)_{\{n\}}$, 
while in  Eq.(19)  $R_i-M_i$ 
follows from   the second relation in Eq.(12).  
The   $C_V$ is   the isochoric heat 
capacity per unit volume. 
Then, $e$ obeys the differential equations,  
\bea
&&\hspace{-2cm} de={\sum}_i M_i d{n}_i+C_V {d T} \\
&&\hspace{-15mm} =   {\sum}_i R_i d{n}_i+T\alpha_s {d p}.   
\ena  
where $T\alpha_s=(\p e/\p p)_{\{ n\}}= 
 C_V(\p T/\p p)_{\{ n\}}$ 
(see Eqs.(26) and (27)). These equations 
follow from the fundamental equation $de= Tds+{\sum}_i \mu_i dn_i$. 
From Eqs.(17)-(19) the differences ${\bar H}_i-{M}_i$ and ${\bar H}_i- {R}_i$ 
 are proportional to ${\bar v}_i$   as 
\be 
 {\bar H}_i = M_i + T\beta_V{\bar v}_i=
 R_i + T\beta_s {\bar v}_i.
\en 
 See the definitions of  
 $C_V$, $\alpha_s$, $\beta_V$, and  $\beta_s$ 
in Eq.(26). We can also 
derive $\beta_s-\beta_V= C_V/T\alpha_p$ from Eq.(27), 
so Eq.(22) is consistent with Eq.(19). 

From Eq.(22) the enthalpy density $h=H/V= {\sum}_i {\bar H}_in_i$ 
is  expressed   as    
\be 
h=e+p={\sum}_i M_in_i+ T\beta_V= {\sum}_i R_in_i+ T\beta_s,  
\en 
 in terms of $M_i$ and $R_i$. 
Thus,   ${\bar H}_i$ can also be expressed as  
\bea
&&\hspace{-10mm} {\bar H}_i = M_i+{\bar v}_i
\big(h-{\sum}_j M_j n_j\big) \\ 
&&\hspace{-5mm} =R_i+{\bar v}_i\big(h-{\sum}_j R_j n_j\big).   
\ena   
 Vlugt \cite{Vlugt} derived    Eq.(24) 
and used it  
 to calculate ${\bar H}_i$ for binary mixtures.

\vspace{2mm}
\noindent{\bf 2.3.  Thermodynamic derivatives 
at fixed  $\{ x \}$}\\ 
In addition to $\alpha_p$, $\kappa_T$, and $C_p$ in Eq.(9), 
  we define  the following   derivatives at fixed   $\{ x\}$:  
\bea 
&&\hspace{-1cm}
C_V=T\ppp{ s}{ T}{\{n\}}=\frac{1}{\gamma}C_p,
 ~~~\beta_V= \ppp{ p}{ T}{\{ n\}},
 ~~  \beta_s= \ppp{ p}{ T}{\sigma,\{ x\}},
\nonumber\\ 
&&\hspace{-5mm}
\alpha_s= \frac{1}{n}\ppp{ n}{ T}{\sigma, \{ x\}},
~~\kappa_s= \frac{1}{n}\ppp{ n}{ p}{\sigma,\{x\}}=
\frac{1}{\gamma}\kappa_T. 
\ena 
Here,  $\alpha_s$ is the adiabatic thermal expansivity   
and $\kappa_s$ is the adiabatic compressibility. 
The     $\alpha_s$,  $\beta_s$, and $\kappa_s$   
are the  isentropic derivatives  at fixed $\sigma=s/n$ 
and $\{ x\}$, while  $\gamma=C_p/C_V=\kappa_T/\kappa_s$ is   the specific 
heat ratio. We  list useful thermodynamic identities,       
\bea
&&\hspace{-1cm} 
 \beta_s= \alpha_s/\kappa_s= C_p/T\alpha_p,~~
\beta_V= \alpha_p/\kappa_T= C_V/T\alpha_s ,
\nonumber\\ 
&&\hspace{-1cm}
{\gamma-1}= \gamma{\beta_V}/\beta_s={\alpha_p}/{\alpha_s}
= T\alpha_p^2/\kappa_TC_V,      
\ena   
which   are well known for pure  fluids 
and    still hold for multi-component fluids at fixed  $\{ x\}$. 

\vspace{2mm}
\noindent{\bf 3. Correlation functions }\\ 
\vspace{2mm}
\noindent{\bf 3.1.  Grand-canonical variances  }\\ 
For any   space-dependent fluctuating 
variables  $\hat{\cal A}({\bi r})$ and 
$\hat{\cal B}({\bi r})$, 
we consider the space integral of their equal-time correlation function 
in the bulk region \cite{Onukibook}, 
\be 
 \av{\hat{\cal A}: \hat{\cal B}}= 
\int \hspace{-0.5mm}
d{\bi r}\av{\delta{\hat{\cal A}}({\bi r})
\delta{{\hat{\cal B}}}({\bi 0})}
\en  
where $\delta{\hat{\cal A}}= {\hat{\cal A}}-\av{{\hat{\cal A}}}$, 
$\delta{\hat{\cal B}}= {\hat{\cal B}}-\av{{\hat{\cal B}}}$, 
  $\av{\cdots}$ denotes   the equilibrium 
 average, and the positions $\bi r$ and  $\bi 0$ 
are located  far from the boundaries. 
The correlation $\av{\delta{\hat{\cal A}}({\bi r})
\delta{{\hat{\cal B}}}({\bi 0})}$   is taken in the grand-canonical 
($T$-$\mu$) ensemble at fixed $V$ 
 \cite{Kubo}, so it   
  decays  rapidly if    $r=|{\bi r}|$ much exceeds    the 
molecular correlation lengths.  
 We call $\av{\hat{\cal A}: \hat{\cal B}}$  
 the $T$-$\mu$ variance of  $\hat{\cal A}$ and 
$\hat{\cal B}$. 
In this paper,  space-dependent quantities 
 with a caret are thermally  fluctuating variables.

The local number densities and the local energy density have 
well-defined microscopic expressions in terms of the 
particle positions and momenta, so they are 
written as 
\be 
{\hat n}_i({\bi r})=n_i + \delta{\hat n}_i({\bi r}),~~
{\hat e}({\bi r})=e + \delta{\hat e}({\bi r}),
\en 
where $n_i$ and $e$ are the equilibrium 
averages  and the second terms are 
the fluctuating deviations. 
 Then,  as will be shown in Appendix A, the thermodynamic derivatives 
  $I_{ij}$ in Eq.(13) are  equal to   the  density variances: 
\be 
I_{ij}= 
\av{{\hat n}_i: {\hat n}_j}=  n_i\delta_{ij} + n_i n_j G_{ij}. 
\en 
where   $G_{ij}$ are   the Kirkwood-Buff (KB) integrals 
\cite{Buff,Oc,Naim2,Naim,Naim3}. 
On the other hand,  
the space integrals of 
the direct correlation  functions $c_{ij}(r)$ 
are equal to $\delta_{ij}/n_i - I^{ij}$.

In this paper, we   consider   the integrals of 
 the  grand-canonical 
 correlation functions of the energy density and the number 
densities: 
\be 
n_iJ_i=  \av{{\hat e}: {\hat n}_i}= \int d{\bi r} 
  \av{\delta{\hat e}({\bi r})\delta {\hat n}_i({\bi 0})}, 
\en 
where  $J_i$ remains finite as $n_i\to 0$. 
Then,  $n_iJ_i$ 
and  $\av{{\hat e}:{\hat e}}$ 
are expressed  as  
\bea 
&&\hspace{-19mm} 
n_i J_i={\sum}_j M_jI_{ji} \\
&&\hspace{-12mm}
={\sum}_j R_jI_{ji} + k_BT^2 \alpha_s n_i\\
&&\hspace{-12mm}
={\sum}_j {\bar H}_jI_{ji} -k_BT^2\alpha_p n_i,\\
&&\hspace{-20mm}
 \av{{\hat e}:{\hat e}}=k_BT^2C_V+{\sum}_{i,j}M_iM_j I_{ij},
\ena
in terms of $M_i$, $R_i$, and ${\bar H}_i$. 
Here,  Eqs.(32) and (35) will be derived 
  in Appendix A, while   
 Eqs.(33) and (34) follow from Eq.(32) 
with the aid of Eqs.(22) and (27).   
Note that Eq.(32) is rewritten as 
 $M_i= {\sum}_j n_j J_jI^{ji}$, 
which was derived by 
Karavias and  Myers \cite{Myers} in their theory 
on heats of adsorption. In Eq.(35),  
the second term is rewritten as ${\sum}_{i,j}\av{{\hat e}: {\hat n}_i}
\av{{\hat e}: {\hat n}_j}I^{ij}$, so $C_V$ is  expressed 
in terms of the $T$-$\mu$ variances.    Mishin \cite{Mishin1} 
obtained an expression equivalent to Eq.(35) 
for the $T$-$\mu$ ensemble.

From Eq.(34) ${\bar H}_i$  are  written  in terms of 
   $J_i$ and  $I_{ij}$ as  
\bea 
&&\hspace{-1cm}
{\bar H}_i=T\beta_V{\bar v}_i+ {\sum}_j n_j J_j  I^{ji}
  \nonumber\\
&&\hspace{-5mm}
= h{\bar v}_i+ {\sum}_j n_jJ_j \big[ I^{ji}-
{{\bar v}_j{\bar v}_i}/{k_BT\kappa_T}\big]  .       
\ena 
where  the second line follows from  Eqs.(14) and (23). 
The first line of Eq.(36)   further  yields 
\be 
  T\alpha_p=T\beta_V\kappa_T=  h\kappa_T  
 -{\sum}_i n_i {\bar v}_iJ_i/k_BT, 
\en  
where  $\alpha_p$ is   written  in terms of the 
$T$-$\mu$ variances   as well as $\kappa_T$ in Eq.(15).   
See simpler expressions of  $\alpha_p$, $\kappa_T$, and $C_p$ 
in Eqs.(67), (A.15),  and (A16) for binary mixtures.

\vspace{2mm}
\noindent
{\bf 3.2. Fluctuations of volume fraction and concentrations}\\
We   introduce   the fluctuations of 
the volume fraction and the concentrations as   
\bea 
&&\hspace{-1cm} 
\delta{\hat\phi}({\bi r})= {\sum}_i {\bar v}_i {\hat n}_i ({\bi r})
-1=   {\sum}_i {\bar v}_i \delta{\hat n}_i ({\bi r}),\\  
&& \hspace{-1cm} 
\delta{\hat X}_i ({\bi r}) =\delta{\hat n}_i({\bi r})/n -
{n_i} \delta{\hat n} ({\bi r})/n^2, 
\ena
where $n={\sum}_i n_i$ and  ${\sum}_i \delta{\hat X}_i=0$. 
See another expression of  $\delta{\hat\phi}$ in Eq.(47). 
 From Eqs.(14), (15), and (37) we find simple variance relations,  
\bea 
&&\hspace{-10mm} \av{{\hat\phi}:{\hat\phi}}= k_BT\kappa_T, ~~
\av{{\hat\phi}:{\hat n}_i}= k_BT\kappa_Tn_i, \nonumber\\
&&\hspace{-10mm}\av{{\hat\phi}:{\hat X}_i}= 0,~~
\av{{\hat\phi}:{\hat e}}=k_BT\kappa_T
(h-T\beta_V) .
\ena
The $\delta{\hat\phi}$ and $\delta{\hat X}_i$ 
 were used originally  by Kirkwood and Goldberg \cite{Kirk3} 
and recently  by our group \cite{Oka1}
in calculating the    scattering   intensities 
(see Eq.(72)).  

The fluctuation  $\delta{\hat\phi}({\bi r})$ is small 
in  nearly incompressible fluids with $\epsilon_{\rm in}\equiv  
k_BTn \kappa_T\ll 1$, where $\epsilon_{\rm in}$ represents 
the degree of compressibility.   
For example, $\epsilon_{\rm in}\sim   0.05$  
for water-alcohol mixtures in the ambient 
condition (at $T \cong  300$ K and $p \cong  1$ atm)  \cite{Kubota}. 

\vspace{2mm}
\noindent
{\bf 3.3. Correlations in the canonical ensemble}\\
In finite systems,  the   correlation functions 
  can have a long-range (nonlocal) part  
in the canonical ensemble \cite{Lebo,Naim1,Kremer,Eg,Ev,Onuki-h}, 
where the  particle numbers $N_i$ and the volume 
$V$ are fixed.  That is, for any  space-dependent fluctuating 
variables ${\hat{\cal A}}({\bi r})$ and ${\hat{\cal B}}({\bi r})$, 
their space correlation in the canonical 
ensemble $\av{\cdots}_{\rm ca}$,  
and that in the $T$-$\mu$ one   $\av{\cdots}$ are related by \cite{Lebo}    
\be 
\av{\delta{\hat{\cal A}}({\bi r})
\delta{{\hat{\cal B}}}({\bi 0})}_{\rm ca}= 
\av{\delta{\hat{\cal A}}({\bi r})
\delta{{\hat{\cal B}}}({\bi 0})}
-  {\sum}_{i,j} D^{ij}_{{\cal A}{\cal B}}I_{ij}/V, 
\en 
where  ${\bi r}$ and  ${\bi 0}$ 
are located  far from the  boundaries.  
In the right hand side, 
the first term  is the  short-ranged 
$T$-$\mu$ correlation, while  
 the second   term is a nonlocal constant proportional to  $V^{-1}$ 
and is small for large $V$. The coefficients 
 $D^{ij}_{{\cal A}{\cal B}}$ are defined by   
\be
 D^{ij}_{{\cal A}{\cal B}}= \ppp{\av{\hat{\cal A}}}{n_i}{T,{n_{k\neq i}}}
 \ppp{\av{\hat{\cal B}}}{n_j}{T,{n_{\ell \neq j}}}
={\sum}_{k,\ell} \av{{\hat{\cal A}}:{{\hat{n}}_k}}
\av{{\hat{\cal B}}:{{\hat{n}}_\ell}}I^{ki}I^{\ell j}, 
\en  
where we use  Eq.(A5). 
Then,  integrating  Eq.(41)     in the cell, we find    
\be 
\frac{1}{V} 
\int_V d{\bi r}_1\int_V  
 d{\bi r}_2 \av{\delta{\hat{\cal A}}({\bi r}_1)
\delta{{\hat{\cal B}}}({\bi r}_2)}_{\rm ca}= 
\av{{\hat{\cal A}}: 
{{\hat{\cal B}}}}-{\sum}_{i,j} 
 \av{{\hat{\cal A}}:{{\hat{n}}_i}}
\av{{\hat{\cal B}}:{{\hat{n}}_j}}I^{ij},
\en  
where the second term is independent of $V$ 
and can be      comparable to the first. 
Note that  the finite-size correction 
due to the boundaries is of order $L^{-1}$ in Eq.(43), 
where $L=V^{1/3}$ is the linear dimension of the cell.  
A simple derivation of Eq.(41) 
was given in Appendix C in   Ref.\cite{Onuki-h}. 
More results on  the canonical correlations will be given 
around  Eq.(52). The nonlocal correlations in the $T$-$p$ 
ensemble will be studied in Section 4.5. 

For $\delta{\hat n}_i$ and $\delta{\hat e}$ 
we rewrite  Eq.(41)  as  
\bea 
&&\hspace{-11mm} \av{\delta{\hat n}_i({\bi r})
\delta{\hat n}_j({\bi 0})}_{\rm ca}=
 \av{\delta{\hat n}_i({\bi r})
\delta{\hat n}_j({\bi 0})} - I_{ij}/V, \nonumber\\
&&\hspace{-11mm}\av{\delta{\hat e}({\bi r})
\delta{\hat n}_i({\bi 0})}_{\rm ca}= 
\av{\delta{\hat e}({\bi r})
\delta{\hat n}_i({\bi 0})} - n_i J_i/V, \nonumber\\
&&\hspace{-11mm}\av{\delta{\hat e}({\bi r})
\delta{\hat e}({\bi 0})}_{\rm ca}= 
\av{\delta{\hat e}({\bi r})
\delta{\hat e}({\bi 0})} - {\sum}_{i,j} M_i  M_j I_{ij}/V, 
\ena 
where we use    Eqs.(18) and (32). 
Here, the space integrals of the 
first two relations vanish and 
that of the third one is 
 $k_BT^2C_V$ from Eq.(35), which   should hold     
in  the canonical ensemble.

\vspace{2mm}
\noindent
{\bf  4. Fluctuations of hydrodynamic variables}\\
\vspace{2mm}
\noindent
{\bf 4.1. Fluctuations of temperature   and pressure}\\
Using  the thermodynamic equations (20) and (21), 
  we  define the local fluctuations  of the temperature and 
the pressure  as {\it linear combinations} 
of $\delta{\hat e}({\bi r})$ and $\delta{\hat n}_i({\bi r})$ 
\cite{Onukibook}:  
\bea 
&&\hspace{-17mm}\delta {\hat T}({\bi r})=\big[\delta{\hat e}({\bi r})
-{\sum}_i M_i \delta{\hat n}_i({\bi r}) \big]/C_V,\\ 
&&\hspace{-17mm}\delta {\hat p}({\bi r})=\big[\delta{\hat e}({\bi r})
-{\sum}_i R_i \delta{\hat n}_i({\bi r}) \big]/T\alpha_s ,
\ena 
where $M_i$  and $R_i$ are defined in Eqs.(18) and (19). 
These  fluctuations  are superimposed  on the 
thermodynamic temperature and pressure $T$ and $p$.  
They consist of the Fourier components 
with   wavelengths longer  than 
the molecular correlation lengths, so they 
obey the Gaussian distributions in the linear regime. These definitions 
  can be used for any statistical  ensembles. 
We remove  $\delta{\hat e}$   from Eqs.(45) and (46) 
 using   Eqs.(19) and (27) to find    
\be 
\delta{\hat\phi}({\bi r})={\sum}_i {\bar v}_i \delta{\hat n}_i({\bi r})
=- \alpha_p \delta{\hat T}({\bi r})
 +\kappa_T \delta {\hat p}({\bi r}) .  
\en 
where $\delta{\hat\phi}({\bi r})$ is given  in Eq.(38).  
Notice that the above fluctuation  equation    assumes  the same form as 
 the thermodynamic differential equation (11).
  
From Eqs.(32)-(35)  and  (40) 
we find  the $T$-$\mu$ variances of $\delta {\hat T}$ and 
$\delta {\hat p}$:     
\bea 
&& \hspace{-1cm}
\av{{{\hat n}_i}:{\hat T}}=0, ~~
\av{{\hat n}_i:{\hat p}}=k_BTn_i, ~~
\av{{\hat e}:{\hat T}}=k_BT^2,~~
\av{{\hat e}:{\hat p}}=k_BT h, 
\nonumber \\
&& \hspace{-1cm}
\av{{\hat T}:{\hat T}}= k_BT^2/C_V, ~~
\av{{{\hat p}}:{\hat T}}=k_BT/\alpha_s,~~
 \av{{\hat p}:{\hat p}}= k_BT/\kappa_s,~~ 
\nonumber \\  
&&\hspace{-1cm} 
\av{{{\hat T}}:{\hat \phi}}=
 \av{{\hat T}:{\hat X}_i}= 0,~~
\av{{{\hat p}}:{\hat \phi}}=k_BT, ~~
 \av{{\hat p}:{\hat X}_i}= 0,
\ena 
The first relation shows that $\delta{\hat T}$ is  {\it orthogonal} to 
$\delta{\hat n}_i$, so Eqs.(45) and (48) yield Eq.(35). 
In Eq.(A12) in Appendix A, we will also introduce 
the entropy fluctuation. 

 In dynamics, the Fourier components of 
the fluctuating hydrodynamic 
 variables have been assumed to obey the linearized hydrodynamic 
equations with random source terms  (as Langevin equations) 
 \cite{Martin,Mountain1,Bernebook},  
where the wavelengths are  longer   than 
the  molecular correlation lengths. 
Their    time correlation functions have been   calculated with 
the  Laplace transformation method  in the time range $t>0$, 
where   their  initial values 
at $t=0$ are given  by  the  variances in  Eq.(48). 
 These results are   used to interpret 
the  dynamic light scattering. 

In the vicinity of  the fluid criticality, 
we can define $\delta {\hat T}$ 
and $\delta{\hat p}$ by    Eqs.(45) and (46)  
 if  the upper cutoff wave number $\Lambda$  
of the fluctuations is   smaller than the inverse 
 correlation length $\xi^{-1}$ (after the renormalization procedure), 
where $\xi$ grows and 
 the most singular  parts of  $\delta{\hat e}$ 
and $\delta{\hat n}_i$ are cancelled 
to vanish. We can further define them  for  $\xi^{-1}\ls 
\Lambda\ll a^{-1}$ 
with $a$ being the typical molecular radius 
  if we assume    a mapping relationship between 
fluids and Ising  ferromagnets \cite{Onukibook,OnukiL,Ani}. 
Indeed, in  near-critical   ferromagnets, 
 the thermal fluctuations of the temperature and the magnetic 
field can be  equated   to 
  the functional derivatives of the (fluctuating) 
Ginzburg-Landau-Wilson 
free energy   (with respect to the energy density 
and the magnetization)  \cite{Hal,Onukibook}, 
 where they  depend on    the magnetization fluctuation  
  nonlinearly  for  $\xi^{-1}\ls \Lambda\ll a^{-1}$ 
(in the course of the renormalization).  

Our  definitions of  $\delta {\hat T}({\bi r})$ 
and $\delta {\hat p}({\bi r})$ in Eqs.(45) and (46) 
are natural   and  rather obvious. It is worth noting that  
Kittel \cite{Kittel} strictly  wrote ''Temperature 
fluctuation is an oxymoron 
because the consistent and consensual 
definition of temperature admits no fluctuation.''

\vspace{2mm}
\noindent
{\bf 4.2.  Entropy deviation in second  order}\\
Greene and Callen \cite{Callen,Callen1} 
obtained    the  correlation moments  of 
general fluctuating   extensive variables by calculating the  
  entropy change induced by them. 
In the same spirit,     Boltzmann's    principle of equal probability 
indicates   that the equilibrium distribution of $\delta{\hat e}({\bi r})$ 
and $\delta{\hat n}_i({\bi r}) $ at long wavelengths is given by   
\be 
P_{\rm eq}(\delta{\hat e}, \delta{\hat n}) = {\rm const.}
 \exp\big[ \int_V d{\bi r} 
(\delta{\hat s})_2({\bi r})/k_B\big], 
\en 
where      $(\delta{\hat s})_2({\bi r})$ is 
the second-order deviation at position $\bi r$ in  the fluctuating 
entropy density in the Gaussian approximation at long wavelengths:     
\bea  
&&\hspace{-15mm} (\delta{\hat s})_2 = 
s({\hat e},\{{\hat n} \})-  
s({ e},\{{ n} \}) -
\delta{\hat e}/T+ {\sum}_i \mu_i\delta{\hat n}_i/T \nonumber\\
&& \hspace{-5mm}
= - {C_V}(\delta {\hat T}/T)^2/{2}  
- k_B{\sum}_{i,j} {I^{ij}} \delta{\hat n}_i \delta{\hat n}_j/2 . 
\ena  
Here,  we perform  the expansion of $s({\hat e},\{{\hat n} \})$ 
with respect to $\delta{\hat e}={\hat e}-{e}$ and 
$\delta{\hat n}_i={\hat n}_i-{ n}_i$, where   $s({e},\{{ n} \})$ is  
the thermodynamic entropy density and 
   $\delta{\hat T}$ is defined by Eq.(45) \cite{Landau-s}.  
Thus,   the above   entropic distribution  
$P_{\rm eq}$ surely  gives the 
$T$-$\mu$  variances in Eq.(48).

\vspace{2mm}
\noindent
{\bf 4.3. Fluctuations of stress tensor and   pressure}\\
The space-dependent  
 stress tensor  ${\hat \Pi}_{\alpha\beta}({\bi r})$ ($\alpha,\beta=x,y,z$) 
has a  microscopic expression 
in terms of the particle  positions and  momenta  
  \cite{Kirk1,Onukibook}. 
The divergence   $-{\sum}_\beta 
\p {\hat \Pi}_{\alpha\beta}/\p x_\beta$  
is the force density on the fluid 
and is equal to  the time derivative of  the momentum 
density (without applied force field). 
The  thermodynamic pressure $p$ is obtained from 
the equilibrium average $\av{{\hat \Pi}_{\alpha\beta}}=\delta_{\alpha\beta} p$ 
in the bulk region far from the boundaries.   

 In  Appendix A, we will  derive variance relations,  
\be
\av{{\hat \Pi}_{\alpha\beta}: {\hat n}_i}= 
\delta_{\alpha\beta} k_BTn_i , ~~ 
\av{{\hat \Pi}_{\alpha\beta}: {\hat e}}= 
\delta_{\alpha\beta} k_BTh, 
\en 
where   the  right hand sides  
of these two relations 
 coincide with  $\delta_{\alpha\beta} 
\av{{\hat p}: {\hat n}_i}$ and $\delta_{\alpha\beta} 
\av{{\hat p}: {\hat e}}$, respectively, from Eq.(48).   
Thus,  the pressure fluctuation  $\delta {\hat p}$ in  
  Eq.(46) is  the {\it  projection} 
of the diagonal part of ${\hat \Pi}_{\alpha\beta}- \delta_{\alpha\beta}p$  
onto $\delta{\hat e}$ and $\delta{\hat n}_i$ \cite{Zwan,Mori,Onukibook}. 
Indeed, if we use   Mori's   linear 
projection operator ${\cal P}$ onto 
the  hydrodynamic fluctuations \cite{Mori},   
we have 
$
{\cal P}{\hat \Pi}_{\alpha\beta}
= \delta_{\alpha\beta}(p+\delta{\hat p}).  
$  
The difference ${\hat \Pi}_{\alpha\beta}^\perp = 
{\hat \Pi}_{\alpha\beta}-\delta_{\alpha\beta}(p+\delta{\hat p})$   
is  {\it orthogonal}   to the hydrodynamic variables,  
which was examined   by some  authors \cite{Wallace,Sasa}.   
In the linear response theory,    
 the    viscosities are 
given by    the  space-time integral 
   of the   dynamical correlation 
$\av{{\hat \Pi}_{\alpha\beta}^\perp({\bi r},t)
{\hat \Pi}_{\gamma\delta}^\perp({\bi 0},0)}$  
 \cite{Onukibook,Mori,Zwan}.  

If ${\cal {\hat B}}={\hat p}$ in Eq.(41), 
 Eqs.(42), (46), and (47)  give   the following canonical correlations:  
\bea 
&&\av{\delta{\hat {\cal A}}({\bi r})\delta{\hat p}({\bi 0})}_{\rm ca}
=\av{\delta{\hat {\cal A}}({\bi r})\delta{\hat p}({\bi 0})}
 -\av{{\hat {\cal A}}:{\hat \phi}}/(V\kappa_T), \nonumber\\
&&\int_V d{\bi r}
\av{\delta{\hat {\cal A}}({\bi r})\delta{\hat p}({\bi 0})}_{\rm ca}
= \beta_V \av{{\hat {\cal A}}:{\hat T}}. 
\ena
 If ${\cal {\hat A}}={\cal {\hat B}}={\hat p}$, 
we have 
$\int_Vd{\bi r}
\av{\delta{\hat {p}}({\bi r})\delta{\hat p}({\bi 0})}_{\rm ca}
 =k_BT(\gamma-1)/\kappa_T$ from 
$D^{ij}_{{p}{ p}}= {\bar v}_i{\bar v}_j/\kappa_T^2$.
However,    $\av{\delta{\hat{\cal A}}({\bi r})\delta{\hat T}({\bi 0})}_{\rm ca}$ has  no nonlocal part  
 for any $\hat{\cal A}({\bi r})$ from $D^{ij}_{{\cal A}{ T}}= 0$.

\vspace{2mm}
\noindent
{\bf 4.4.  Landau-Lifshitz theory   }\\
 Landau and Lifshitz \cite{Landau-s}  
introduced    fluctuations of the   temperature and  the pressure,  
 written as  $\Delta T$ and $\Delta p$, which    
  are linear combinations of   extensive changes  
 in  the  volume and  the energy,    
 $\Delta V$ and   $\Delta E$. 
They  used  the $T$-$p$ distribution 
\cite{Kubo,Binderbook,Fre}, where the  particle numbers 
$N_i$ are  fixed but     the  volume fluctuation 
$\Delta V$  is superimposed on   the average 
volume $V$ with the $T$-$p$ variances \cite{Landau-s} given by   
\bea 
&&\hspace{-10mm}\av{(\Delta V)^2}
= Vk_BT \kappa_T,~~\av{\Delta V\Delta p}
=-k_BT, \nonumber\\
&&\hspace{-12mm}
\av{\Delta V\Delta T}
= 0,~~\av{\Delta V\Delta E}= Vk_BT \kappa_T(T\beta_V-p).  
\ena 
These relations hold for multi-component fluids, where 
  use  is made of the thermodynamic derivatives at fixed $\{ x\}$ 
in Eqs.(9) and (26).  Here, we note that 
   $\Delta T$ and $\Delta p$  
in the   Landau-Lifshitz book   
are the space averages of  $\delta {\hat T}({\bi r})$ and 
$\delta {\hat p} ({\bi r})$ in Eqs.(45) and (46):    
\bea 
\Delta T=\frac{1}{V} \int_V d{\bi r}\delta {\hat T}({\bi r}),~~ 
\Delta p=\frac{1}{V} \int_V d{\bi r}\delta {\hat p}({\bi r}).
\ena 
We  can define $\Delta T$ and $\Delta p$ 
by Eq.(54)  for any ensembles. 

With  nonvanishing   $\Delta V$, 
 the particle-number changes 
 $\Delta N_i$ and the internal-energy change 
$\Delta E$  are expressed   in the linear order as  
\be 
\Delta N_i= \int_V d{\bi r} 
\delta{\hat n}_i+ n_i\Delta V,~~
 \Delta E= \int_V d{\bi r}\delta {\hat e}+ e\Delta V, 
\en 
With the aid of    Eqs.(23), (45),  and (46)  
we can  then express   $\Delta T$ and $\Delta p$  as 
\bea 
&&\hspace{-1cm} 
\Delta T= \Big[\Delta E+ (p-T\beta_V)\Delta V
-{\sum}_i M_i \Delta N_i\Big]/VC_V,\nonumber\\
&&\hspace{-1cm} 
\Delta p =\Big[\Delta E+ (p-T\beta_s)\Delta V
-{\sum}_i R_i \Delta N_i\Big]/(VT\alpha_s). 
\ena 
If $\Delta N_i=0$, these  expressions  are 
equivalent to those   in the Landau-Lifshitz book. 

In   the  $T$-$\mu$,  canonical, and $T$-$p$  ensembles, 
 $\Delta T$  commonly satisfies the  relations,  
\be 
 \av{(\Delta T)^2}= { k_BT^2}/{VC_V}, ~~
 \av{\Delta T\Delta E}=  {k_BT^2},   
\en  
which follow  from   Eqs.(48) and (55). 
The fluctuation of $\Delta T$ is  small for large $V$. 
We  mention   an experiment \cite{Lipa} 
and   theories \cite{Long,Schm,Mishin} on $\Delta T$ 
in a small system in contact with a heat bath.

\vspace{2mm}
\noindent
{\bf 4.5. Correlations in the $T$-$p$ ensemble}\\
Simulations in the $T$-$p$ 
ensemble have been performed extensively
 \cite{Fre,Binderbook}, which are powerful 
for investigating phase coexistence 
at fixed $T$ and $p$.  
Since the finite-size effects are essential 
in   any ensembles \cite{Bedeaux,Kremer}, 
 we here present  the  counterpart of Eq.(41) 
for   the $T$-$p$ ensemble.    
Namely, for any space-dependent fluctuating variables 
 ${\hat{\cal A}}({\bi r})$ and ${\hat{\cal B}}({\bi r})$, 
we relate 
their   $T$-$p$ correlation $\av{\cdots}_{Tp}$, 
   $T$-$\mu$ one   $\av{\cdots}$, and   canonical one 
  $\av{\cdots}_{\rm ca}$ as 
\bea 
&&\hspace{-24mm} 
\av{\delta{\hat{\cal A}}({\bi r})
\delta{{\hat{\cal B}}}({\bi 0})}_{Tp}= 
\av{\delta{\hat{\cal A}}({\bi r})
\delta{{\hat{\cal B}}}({\bi 0})}
-  {\sum}_{i,j} D^{ij}_{{\cal A}{\cal B}}
\big[I_{ij}- k_BT\kappa_T n_in_j\big]/V\nonumber\\
&&\hspace{6mm} =\av{\delta{\hat{\cal A}}({\bi r})
\delta{{\hat{\cal B}}}({\bi 0})}_{\rm ca}
+ \av{{\hat{\cal A}}:{{\hat{\phi}}}}
\av{{\hat{\cal B}}:{{\hat{\phi}}}}/(k_BT\kappa_TV), 
\ena 
where we define  $D^{ij}_{{\cal A}{\cal B}}$ in Eq.(42) 
and $\delta{\hat\phi}$ in Eq.(38) and use Eq.(14). 
Here, $I_{ij}= \av{{\hat n}_i: {\hat n}_j}$ in  Eq.(41) 
are replaced by  $\av{{\hat n}_i': {\hat n}_j'}=
I_{ij}- k_BT\kappa_T n_in_j$ with  
  $\delta{\hat n}_k'=\delta{\hat n}_k- n_k \delta{\hat \phi}$,   
where     $n_k \delta{\hat \phi}$ are        
the dilational density changes and  
 $\delta{\hat n}_k'$ vanish  at fixed concentrations. 
From $D^{ij}_{{\cal A}{\cal T}}=0$, 
the   correlation of $\delta{\hat T}({\bi r})$ and 
 any $\delta{\hat{\cal A}}({\bi r})$ has  no nonlocal part  
in the three  ensembles.

With the aid of   Eqs.(23), (32), and (35),    Eq.(58) yields    
\bea 
&&\hspace{-4mm}
\int_V\hspace{-1mm} d{\bi r} \av{\delta{\hat{ n}_i}({\bi r})
\delta{{\hat{n}_j}}({\bi 0})}_{Tp}/k_BT
= \kappa_Tn_in_j,\nonumber\\
&&\hspace{-4mm}
\int_V \hspace{-1mm}d{\bi r} \av{\delta{\hat{ e}}({\bi r})
\delta{{\hat{n}_i}}({\bi 0})}_{Tp}/k_BT =  
 (h- T\beta_V)\kappa_Tn_i,\nonumber\\
&&\hspace{-10mm}
\int_V \hspace{-1mm}
d{\bi r} \av{\delta{\hat{ e}}({\bi r})\delta{{\hat{e}}}({\bi 0})}_{Tp}
/k_BT= TC_V+ (h- T\beta_V)^2\kappa_T. 
\ena  
We  can then derive  Eq.(53) using Eq.(55). Here,   
with  replacement  $h-T\beta_V \to p-T\beta_V$,  
 the right hand side of  the third relation   
becomes equal to  $\av{(\Delta E)^2}_{Tp}/V$ 
\cite{Landau-s,Kubo,Mishin1}. 
 Also    $\av{(\Delta p)^2}V$ is given by 
$k_BT/\kappa_s$ in the $T$-$\mu$ and $T$-$p$ 
ensembles  and  by  
$k_BT(1/\kappa_s-1/\kappa_T)$ in the canonical one 
from Eqs.(48),  (58), and (52), respectively.

For one-component fluids, the second  term in Eq.(58) 
vanishes, so there is   no nonlocal correlation 
 in  the $T$-$p$ ensemble,   which is in accord with 
a previous simulation \cite{Velasco}. For binary mixtures, 
the nonlocal part in  Eq.(58)  is rewritten  as 
\be 
\av{\delta{\hat{\cal A}}({\bi r})
\delta{{\hat{\cal B}}}({\bi 0})}_{Tp}-  
\av{\delta{\hat{\cal A}}({\bi r})
\delta{{\hat{\cal B}}}({\bi 0})}= 
- \ppp{\av{\hat{\cal A}}}{ X}{T,p} 
\ppp{\av{{\hat{\cal B}}}}{ X}{T,p} \frac{\chi}{V}  
= - \frac{\av{{\hat{\cal A}}: {\hat X}}
\av{{\hat{\cal B}}: {\hat X}}}{\av{{\hat{ X}}: {\hat X}}V},
\en 
where   $\delta{\hat X}$ is 
 the concentration fluctuation   in Eq.(61)  with 
$\chi= \av{{\hat{ X}}: {\hat X}}$ in Eq.(63).    
We here  use Eq.(62), the relations below Eq.(70), and 
Eq.(A14).

\vspace{2mm}
\noindent
{\bf 5. Binary mixtures} \\
\vspace{2mm}
\noindent
{\bf 5.1. Partial quantities and correlations} \\
For binary mixtures, the fluctuating   concentration 
 is written as ${\hat X}=X+\delta {\hat X}$ with  $X= n_2/n$ 
and $\delta{\hat X}({\bi r})=\delta {\hat X}_2({\bi r})$  
from Eq.(39). From Eqs.(38) and (39)  we have   
\be 
\delta {\hat \phi}= {\bar v}_1\delta {\hat n}_1
+  {\bar v}_2\delta {\hat n}_2,~~  
\delta {\hat X}= n^{-2}( n_1\delta {\hat n}_2- n_2\delta {\hat n}_1). 
\en 
where   $n=n_1+n_2$. 
Then,  $\delta {\hat n}_1$ and $\delta {\hat n}_2$ are 
written as 
\be 
\delta {\hat n}_1= n_1 \delta {\hat \phi}-{\bar v}_2n^2 
\delta {\hat X}, ~~\delta {\hat n}_2= n_2 \delta {\hat \phi}+{\bar v}_1n^2 
\delta {\hat X}. 
\en 
Since  $\av{{\hat \phi}: {\hat X}}= 0$ from  Eq.(40), 
$I_{ij}$ are expressed in terms  of 
the compressibility  $\kappa_T= \av{{\hat \phi}: {\hat \phi}}/k_BT$ 
and the concentration susceptibility 
 $\chi$ with    
\be
\chi= \av{{\hat X}: {\hat X}}=\frac{k_BT}{n} \ppp{X}{ \Delta}{T,p} 
 =\frac{n_1n_2}{n^3}+ \frac{n_1^2n_2^2}{n^4} 
 (G_{11}+G_{22}-2G_{12}) ,   
\en 
where   $\Delta=\mu_2-\mu_1$ 
is the chemical potential difference 
(see Appendix A). We have $\chi\cong {n_1n_2}/{n^3}$ 
for dilute  mixtures or 
for mixtures composed of {\it similar} components 
away from the criticality.   
However, $\chi$ divereges at the  criticality 
(see Section 5.3). 
The spinodal condition   $\chi=\infty$  
 was    examined theoretically 
 for highly asymmetric, hard-sphere  mixtures  
 \cite{Biben,Lekker}.

In the original papers \cite{Buff,Naim2,Naim,Oc}, 
the partial volumes ${\bar v}_1$ and ${\bar v}_2$ 
are expressed in terms of $G_{ij}$ from Eq.(14). 
They are also expressed in terms of $\hat X$ and $\chi$ as    
\bea 
&&\hspace{-14mm}
{\bar v}_1= k_BT\kappa_T [I_{22}n_1-I_{12}n_2]/D= 
\av{{\hat n}_2: {\hat X}}/n^2\chi, \nonumber\\
&&\hspace{-14mm}{\bar v}_2=k_BT\kappa_T 
[I_{11}n_2-I_{12}n_1]/D= 
 -\av{{\hat n}_1: {\hat X}}/n^2\chi,  
\ena 
where $D$ is  the determinant of $\{ I_{ij}\}$ 
and  is calculated from Eq.(62) as  
\be 
D=I_{11}I_{22}- I_{12}^2= n^4k_BT\kappa_T\chi.
\en 
In the dilute limit of the second component,  Eq.(64) yields \cite{Buff}  
\be 
 {\bar v}_2^0= \lim_{n_2\to 0}{\bar v}_2= 
 \lim_{n_2\to 0}(k_BT\kappa_T - G_{12}).   
\en 
In addition, if use is made of  ${\hat n}={\hat n}_1+ {\hat n}_2$, 
 Eq.(65)  is rewritten as 
\be 
n^2k_BT\kappa_T
= \av{{\hat n}: {\hat n}}- 
\av{{\hat n}: {\hat X}}^2/\av{{\hat X}: {\hat X}}.  
\en  
See similar expressions for $\alpha_p$ and $C_p$ in 
Eqs.(A15) and (A16). 

The partial enthalpies 
 ${\bar H}_i$ in Eq.(36) are expressed as 
\bea 
&&\hspace{-12mm}
{\bar H}_1=h{\bar v}_1- \av{{\hat e}:{\hat X}}\frac{n_2}{n^2\chi} 
=h{\bar v}_1- \ppp{e}{X}{T,p}\frac{n_2}{n^2}, \nonumber\\
&&\hspace{-12mm}
{\bar H}_2=h{\bar v}_2+ \av{{\hat e}:{\hat X}}\frac{n_1}{n^2\chi}
=h{\bar v}_2+\ppp{e}{X}{T,p}\frac{n_1}{n^2}.   
\ena 
In terms of  $J_i$ in Eq.(31) 
we  express  $\av{{\hat e}:{\hat X}}$  as 
\be 
\av{{\hat e}:{\hat X}}=
(J_2-J_1)n_1n_2/n^2.
\en 
From Eqs.(A14) and (34)  we  find  
\be 
 \ppp{e}{X}{T,p}= 
\av{{\hat e}:{\hat X}}/\chi =
({\bar v}_1 {\bar H}_2 
-{\bar v}_2 {\bar H}_1 )n^2,  
\en 
which  also follows from 
 Eqs.(10) and (11). That is,  at fixed $(T, p)$, we have      
$dn_1= -n^2{\bar v}_2 dX$,  $dn_2= n^2{\bar v}_1dX$, 
and $de= dh= {\bar H}_1 dn_1+ {\bar H}_2 dn_2$, 
so $(\p w/\p X)_{T,p}/n^2= {\bar v}_1(\p w/\p n_2)_{T,n_1}- 
{\bar v}_2(\p w/\p n_1)_{T,n_2}$ for any $w$.   
From   Eqs.(68) and (69) 
 the infinite-dilution limit 
of ${\bar H}_2$ becomes 
\be
 {\bar H}_2^0=\lim_{n_2\to 0}
{\bar H}_2 =\lim_{n_2\to 0}(h{\bar v}_2+ J_2-J_1).
\en

The correlation-function 
expressions  (64) and (68) are convenient 
for numerical calculations of 
   ${\bar v}_i$ and  ${\bar H}_i$. They also indicate how 
   ${\bar v}_i$ and  ${\bar H}_i$ 
behave near the fluid criticality (see the next subsection). 
We mention  a review on  the 
calculations of ${\bar H}_i$ in Ref.\cite{Vlugt}.

\vspace{2mm}
\noindent
{\bf 5.2. Scattering intensity and  alcohol-water mixtures} \\
In scattering experiments from   binary mixtures,  
the measured intensity   in the long wavelength limit is given by 
 $I= \av{\hat{\cal C}:\hat{\cal C}}$ with  
 $\delta\hat{\cal C}({\bi r})= b_1\delta{\hat n}_1+ b_2\delta{\hat n}_2
+b_T \delta {\hat T}$, where 
  $b_1$,  $b_2$, and $b_T$  are  constants. We 
express $\delta{\hat n}_1$ and $\delta{\hat n}_2$ 
 in terms of $\delta{\hat \phi}$ and $\delta{\hat X}$ using 
 Eq.(62) to   find 
\be 
 I =A_\phi^2k_BT\kappa_T 
+A_X^2\chi  + b_T^2 k_BT^2/C_V ,
\en 
where $A_\phi={\sum}_i b_in_i$
and $A_X=n^2(b_2{\bar v}_1- b_1{\bar v}_2)$. 
In their  theory on  light scattering, 
Kirkwood and Goldberg \cite{Kirk3} set 
 $A_\phi=n(\p\epsilon/\p n)_{T,X}$ 
and   $A_X=(\p\epsilon/\p X)_{T,p}$ using  
  the optical dielectric constant   $\epsilon$ \cite{Kirk3}.  
 Bhatia and  Thornton \cite{Bhatia} calculated  $I$ using  
$\delta{\hat n}=\delta {\hat n}_1+\delta {\hat n}_2$ 
and $\delta{\hat X}$ with a 
 cross term $(\propto \av{{\hat n}:{\hat X}}$). 
In these theories, the last term in Eq.(72) 
due to $\delta{\hat T}$  was  neglected.

In scattering experiments on 
 alcohol-water mixtures in the ambient condition, 
the concentration susceptibility 
$\chi$ in Eq.(63) exhibited  a maximum     at intermediate $X$,  
where  $\chi\gs 1/n$ and $G_{11}+G_{22}-2G_{12}\gg 1/n$   
\cite{Oka1,Nishi,Misawa,YKoga,Bagchi1,Lep,Ruck,Koga}, 
This is because 
 an amphiphilic species   (such as  ethanol 
and  tertbutyl-alcohol)  tends to 
    aggregate in  liquid water  on nanometer scales,

\vspace{2mm}
\noindent
{\bf 5.3. Critical behaviors in binary mixtures  } \\
 In  near-critical  binary mixtures 
\cite{Gri,Wh,Cum,LSengers,Chimo,Eck1,Eckert,Onukibook,OnukiH}, 
  ${\bar v}_i$ and  ${{\bar H}_i}$ 
 exhibit  complicated  crossover  behaviors, which has not been 
  studied.  Near the gas-liquid criticality   
of the  primary component  1, ${\bar v}_2^0$ and  ${{\bar H}_2^0}$ 
at infinite dilution $n_2\to 0$ grow strongly from     Eqs.(66) and (71). 
Experimentally,   $v_2^0$ became negatively large 
near the solvent  gas-liquid critical point 
    \cite{Eckert,Chimo}. 
However,    ${\bar v}_2$ and  ${{\bar H}_2}$ 
tend to constants  for $n_2k_BT\kappa_T \gs 1$ with increasing  
$\kappa_T $ from Eqs.(64) and (68).

 On approaching  a consolute critical line,  
 ${\bar v}_i$ and  ${{\bar H}_i}$ 
tend to  finite values. 
This is because  $\av{\hat{n}_i:{\hat X}}$, $\av{\hat{e}:{\hat X}}$,  
and $\chi$ in Eqs.(64) and (68)    
grow  weakly  near the non-azeotropic 
criticality (with the critical exponent 
${\hat\alpha}\cong 0.10$), but they grow strongly  near the azeotropic 
criticality (with the critical exponent 
${\hat\gamma}\cong 1.24$)  \cite{Gri,Onukibook,Chimo}.

\vspace{2mm}
\noindent {\bf 6. Dilute solute in mixture solvents}\\
\vspace{2mm}
\noindent{\bf 
6.1. Helmholtz free energy up to order $n_3^2$ }\\
We consider a mixture solvent composed of 
two components  $(i=1, 2$)    and   add a dilute 
nonionic solute $(i = 3)$. 
Their densities are written as $n_i$ 
with $n_3\ll  n_1+ n_2$.  The solute 
has no internal degree of freedom  
and there is  no chemical reaction. 
In this  section, the solvent 
composition is represented by the concentration of the second component, 
\be 
X= n_2/(n_1+ n_2).
\en 
In this and next sections, we consider 
the thermodynamic deviations in  the solvent, $\delta n_1$, 
$\delta n_2$,  $\delta\phi= 
{\bar v}_1\delta { n}_1+  {\bar v}_2\delta { n}_2$ 
and $\delta {X}= n^{-2}( n_1\delta { n}_2- n_2\delta { n}_1)$ 
with $n= n_1+ n_2$ (not including $n_3$). 
Note that  $\delta{\hat n}_1$, 
$\delta{\hat n}_2. $$\delta{\hat\phi}$, and $\delta {\hat X}$
in the previous sections  denote the thermal fluctuations.

For small $n_3$ we  set up 
 the Helmholtz free energy density  
  up to order $n_3^2$  \cite{Oka1,OnukiH}:  
\be 
 f(\{ n\},T)= f_m(n_1,n_2,T) 
+ k_BT \big[ \ln (n_3\lambda_3^3)-1+\nu_3\big]n_3 +
\frac{1}{2}k_BT u_3 n_3^2 , 
\en 
where    $ f_m$  is  the  solvent free energy density,  
 and  $\lambda_3 (\propto T^{-1/2})$ is the thermal 
de Broglie length of the solute.  The  dimensionless 
quantity $\nu_3(n_1,n_2, T)$ is crucial  
in this section. It  contains an ideal-gas  part \cite{Landau-s},       
\be 
  \nu_3^0(T)=  -(k_3-3/2)\ln (T/\Theta_3^0) , 
\en 
where  $2k_3-3$ is   the vibration-rotation   
 degree of   freedom of a  solute particle and   
  $\Theta_3^0$ is a  constant temperature. 
The solute contribution to 
the  ideal-gas part of 
$C_V=(\p e/\p T)_{\{n\}}$   is  then   $k_Bk_3n_3$ 
(see Eq.(77)). 
The difference     $\Delta\nu_3= \nu_3-\nu_3^0$ is determined by  
the solute-solvent interactions at infinite dilution. 
The last term in Eq.(74) ($\propto n_3^2$)  arises from  
the {\it direct} solute-solute 
interaction under the influence of the solvent, so $u_3$ can depend on 
$(n_1,n_2,T)$. See the {\it indirect} solute-solute  interaction    
from our model in Eq.(93).

From Eq.(74)  the solute chemical potential $\mu_3= \p f/\p n_3$, 
the  energy density $e=-T^2\p (f/T)/\p T$,  
and the pressure $p= {\sum}_i n_i\p f/\p  n_i-f$ 
are calculated as  
\bea 
&&\hspace{-12mm}
 \mu_3/  k_BT = 
\ln (n_3\lambda_3^3)+\nu_3 + u_{3} n_3, 
\\
&&\hspace{-12mm}
e/k_BT= e_m/k_BT+ \big( \frac{3}{2}-T\nabla_T\nu_{3}\big) n_3 
-\frac{1}{2}{ n_3^2}T\nabla_T {u_{3}} . \\ 
&&\hspace{-12mm}
p/k_BT= p_m/ k_BT + (1+ \zeta_3)n_3+\frac{1}{2}  n_3^2  
\nabla_n ({ n}u_3),  
\ena 
where   $e_m(n_1,n_2,T)$ is the  solvent energy density 
and  $p_m(n_1,n_2,T)$  the solvent pressure.\\ 

We consider the derivatives of $\nu_3(n_1,n_2,T)$. 
Setting   $\nabla_T w= 
(\p w/\p T)_{\{n\}}$ and   $\nabla_n w= 
(\p w/\p n)_{T, X}$ with  $n=n_1+ n_2$ for any  $w$, we have  
\bea 
&&\hspace{-13mm}
\nu_{31}= (\p \nu_3/\p n_1)_{T,n_2},~~
\nu_{32}= (\p \nu_3/\p n_2)_{T,n_1},\\
&&\hspace{-13mm}
\zeta_3= n_1{\nu_{31}} +n_2{\nu_{32}}
= n\nabla_n  \nu_3,\\
&&\hspace{-13mm}
g_3 =  n^2({\bar v}_1 \nu_{32}-{\bar v}_2 \nu_{31})
= (\p \nu_3/\p X)_{T, p}, 
\ena  
In Eq.(81) we can  neglect 
 the $n_3$-dependence of ${\bar v}_1$ and ${\bar v}_2$ 
 and use  the first relation in Eq.(12).  Here,  $\nu_{3i}$, $\zeta_3$, 
and $g_3$  remain finite  in the  limit 
$X\to 0$ and 1 and   will be expressed 
in terms of the KB integrals in Eqs.(98)-(100). 
For small deviations of $n_1$, $n_2$, and $T$, we find   
\be
\delta\nu_3= \nu_{31}\delta n_1+ \nu_{32}\delta n_2
 + (\nabla_T \nu_3)  \delta T
= \zeta_3\delta\phi+ g_3 \delta X+  (\nabla_T \nu_3)  \delta T.
\en
The parameter $g_3$  represents the 
asymmetry between the solute-solvent (3-1) interaction and 
the solute-cosolvent (3-2) one  
(see Eq.(100)). 

We make some remarks regarding the interaction 
part $\Delta\nu_3=\nu_3-\nu_3^0$.\\
 (i) For  ternary solutions,  Ben-Naim  \cite{Naim3} 
introduced a solute {\it psudo-chemical} potential, 
written as $\mu_s^{*\ell}$, 
which is equal to  $k_BT \Delta\nu_3$ in our notation. 
He  expressed the derivative 
$(\p  \mu_s^{*\ell}/\p X)_{T,p}$ 
 in terms of the KB integrals, which 
coincides with  $ g_3$  in Eq.(100) multiplied by $k_BT$.\\ 
(ii) If the solute volume  ${\bar v}_3$ 
 exceeds  $1/n$ in a liquid  solvent, 
 each solute particle is in contact with many 
solvent particles, leading to 
 $|\Delta\nu_3|\gg 1$ for
 not  small preferential solvation. See  numerical calculations  of 
$\Delta\nu_3$, $\zeta_3$, and $g_3$ vs $X$  for various ${\bar v}_3$ 
 in Fig.3 in Ref.\cite{Oka1}. 
For hard-spheres  
in ambient liquid water, Chandler {\it et al.}  \cite{Chan}
concluded that   $\Delta G(= k_BT\Delta\nu_3)$ 
is proportional to $R^2$ 
for  $R\gs 1$ nm with  $R$ being  the sphere radius, 
which yields $\Delta\nu_3\sim 200$ at  $R\sim 1$ nm. More discussions  
will be given around Eqs.(89) and (97).\\ 
(iii) 
When the solution  is phase-separated into liquid 
and gas regions, the Ostwald  coefficient $L$ 
($=$the ratio of the solute densities 
in the two phases)  \cite{Battino}   is given by 
\be 
L= n_3^{\rm liq}/n_3^{\rm gas} 
= \exp[ -\Delta\nu_3],
\en 
in the dilute limit 
of the solute  \cite{Oka1,LSengers,Ben,OnukiH}. 
The solute solubility  increases  with 
decreasing $\Delta\nu_3$, which is the case 
of increasing $X$ for  $g_3<0$ from Eq.(82).

\vspace{2mm}
\noindent{\bf 6.2. Solute partial quantities  in terms of $\nu_3$  
and critical behaviors}\\
Using  Eqs.(76)-(78), 
we  can  calculate ${\bar v}_3$ in  Eq.(12),  
   ${\bar H}_3$ in  Eq.(17), 
and $M_3$ in  Eq.(18)     as 
\bea 
&&\hspace{-17mm}
{\bar v}_3/ k_BT\kappa_T= 
1+ \zeta_3+ n_3  \nabla_n (n u_{3}),  
\\ 
&&\hspace{-18mm}
{\bar H}_3/k_BT  
= {3}/{2} -T\nabla_T \nu_{3} + T\alpha_p(1+\zeta_3) 
  - n_3T\big[\nabla_T u_{3}- \alpha_p 
 \nabla_n (n u_{3})\big], \\
&&\hspace{-18mm} 
  M_3/k_BT=  {3}/{2} -T\nabla_T \nu_{3}-  n_3 \nabla_T u_3 ,
\ena 
which are valid  up to order $n_3$. 
In the infinite-dilution limit  $n_3 \to 0$,   we set 
\bea 
&&\hspace{-5mm}
\kappa_T \to  \kappa_T^m, 
~~\alpha_p \to  \alpha_p^m,~~\beta_V \to  \beta_V^m.~~ 
{\bar v}_3 \to  {\bar v}_3^0, ~~  
 {\bar H}_3 \to {\bar H}_3^0, ~~G_{3i}\to G_{3i}^0 , ~~J_3\to J_3^0, 
\nonumber\\
&&\hspace{-5mm}
{\bar v}_i\to {\bar v}_i^m,~~
  I_{ij} \to I_{ij}^m,  ~~J_i \to J_i^m ~~(i,j=1,2).
\ena
From   Eqs.(84) and (85) we then obtain the infinite-dilution limits of 
 ${\bar v}_3$ and ${\bar H}_3$: 
\bea 
&&\hspace{-15mm}
{\bar v}_3^0=\lim_{n_3\to 0}{\bar v}_3= 
  k_BT\kappa_T^m(1+ \zeta_3),  
\\ 
&&\hspace{-15mm}
{\bar H}^0_3 =\lim_{n_3\to 0}{\bar H}_3
= k_BT(3/2-T\nabla_T\nu_3) +T\beta_V^m {\bar v}_3^0.  
\ena 

For relatively large solutes 
in nearly incompressible solvents with $n{\bar v}_3^0\gs 1$ and 
$nk_BT\kappa_T^m\ll 1$, 
  Eq.(88) indicates $\zeta_3\gg 1$ and   
${\bar v}_3^0\cong  k_BT\kappa_T^m \zeta_3
= k_BT (\p \nu_3/\p p)_{T,X}$.  
For example, if $n{\bar v}_3^0 \sim 5$, 
we have $\zeta_3\sim 100$ in ambient water-alcohol 
mixtures \cite{Kubota}. In   addition,   $\kappa_T$,  $\alpha_p$, 
 ${\bar v}_i$,  and ${\bar H}_i$ 
($i=1,2$) will be expanded up to order $n_3$ in  Appendix B.  

Let us consider a near-critical 
 mixture solvent. 
Here, $\nu_3$, $\nabla_T\nu_3$, and $\zeta_3$  
are regular  functions of $(T, n_1, n_2)$  
even near the  criticality, while  they are  singular 
as functions of $(T,p,X)$ near the criticality. 
As a result, the critical behaviors of 
$ {\bar v}_3^0$ and ${\bar H}_3^0$ are  the same  as 
that of  $\kappa_T^m$   from Eqs.(88) and (89), so  
 they grow weakly (strongly) near the non-azeotropic (azeotropic) criticality 
 (see Section 5.2). 
Note that many binary mixtures composed of 
 two {\it similar} components are {\it nearly azeotropic} 
at any $X$ \cite{Ani,OnukiL}, 
 where the  singular part of $\delta{\hat X}$ is  small 
 and ${\bar v}_3^0$ and  ${{\bar H}_3^0}$ 
exhibit large critical enhancement.

\vspace{2mm}
\noindent
{\bf 6.3. Solute density variances  } \\
Using the second relation in 
Eq.(13) and $\mu_3$ in Eq.(76), we obtain   
\be  
I^{3i}= \nu_{3i}~~ (i=1,2) , ~~I^{33}= 1/n_3 + u_3,
\en 
where we neglect  $n_3\p u_3/\p n_i$  in $I^{3i}$ 
but retain $u_3$  in $I^{33}$ (to obtain Eq.(93) below). From  
 ${\sum}_{k=1,2}I^{ik}I_{k3}= -I^{3i}I_{33}= 
-\nu_{3i}I_{33}$ ($i=1,2)$, 
we have  
\be
I_{3i}/I_{33}=  -{\sum}_{k=1,2} \nu_{3k}I^m_{ik} 
  ~~(i=1,2). 
\en 
Then,    $1-I^{33}I_{33} ={\sum}_{i=1,2}I^{3i}I_{i3}
= -I_{33} {\sum}_{i,j=1,2}
 \nu_{3i}\nu_{3j} I_{ij}^m$ from Eqs.(90) and (91), so we obtain  
\be
I_{33}=[I^{33}+ {\sum}_{i,j=1,2}
 I_{ij}^m \nu_{3i}\nu_{3j}]^{-1}= 
  n_3/ [1+ n_3 u_{3}^{\rm eff}], 
\en  
where    $u_{3}^{\rm eff}$ is the effective solute-solute  interaction 
parameter calculated as  \cite{Oka1} 
\bea 
&&\hspace{-27mm}
u_{3}^{\rm eff}= {u_{3}}- {\sum}_{i,j=1,2} I^m_{ij}\nu_{3i}\nu_{3j}
\\
&&\hspace{-20mm} =u_3-  k_BT\kappa_T^m \zeta_3^2 - \chi g_3^2.
\ena   
 In  Eq.(93)  the second term   
 stems from  the solvent-mediated solute-solute 
 attraction. In  Eq.(94) it  is divided into the 
contribution from the volume-fraction 
change ($\propto \zeta_3^2$) 
 and that from the composition 
change ($\propto g_3^2$), where 
  $\zeta_3$ and  $g_3$   are given 
in  Eqs.(80) and (81)  and $\chi$ 
is  defined in Eq.(63) for binary mixtures. 
From $I_{33}= n_3+ n_3^2G_{33}$ we also have 
\be 
G_{33}= -  u_{3}^{\rm eff}/(1+n_3u_{3}^{\rm eff}), ~~~
G_{33}^0=\lim_{n_3\to 0}G_{33}=  -  u_{3}^{\rm eff}.
\en 
If we  superimpose small  
density deviations $\delta n_i$ 
on $n_i$ at fixed $T$, the second order deviation of 
 the displaced  free energy density $f(T, \{ n+\delta n\})$ is 
given by ${\sum}_{i,j} I^{ij}\delta n_i\delta n_j/2$ from Eq.(13). 
This expansion  
 also leads  to  Eqs.(91) and (92) \cite{Oka1}.

  Let us assume $u_{3}^{\rm eff}<0$ and   
$|u_{3}^{\rm eff}|\gg 1/n$, which are realized 
 with increasing  the solvent-mediated  attraction. 
 In this situation, 
we  can define the  solute   spinodal density $n_3^{\rm sp}$ by   
\be 
I_{33}=n_3/(1-n_3/n_3^{\rm sp}),~~~
n_3^{\rm sp}= -1/u_{3}^{\rm eff}= 1/G_{33}^0 \ll n,
\en 
The solute fluctuation grows  as 
$n_3\to n_3^{\rm sp}$,  indicating 
 solute aggregation  as  $n_3$ aproaches $ n_3^{\rm sp}$. 

We  further make remarks regarding  Eq.(94).\\ 
(i) The  second  term in  Eq.(94) 
 is rewritten   as   
\be 
-k_BT\kappa_T^m \zeta_3^2=  
-({\bar v}_3^0- k_BT\kappa_T^m)^2 /(k_BT\kappa_T^m), 
\en 
For relatively large solutes with 
${\bar v}_3^0\gs 1/n$ in  nearly 
incompressible solvents,  we find  
$k_BT\kappa_T^m \zeta_3^2
\gs 
1/(n^2 k_BT\kappa_T^m \gg 1/n$ (see the paragraph below Eq.(89)). 
Thus,     large  attraction arises   
among large solute particles, which 
is the case even for one-component solvents or 
in the limit $X\to 0$ for binary mixtures.   
This explains the   hydrophobic 
  aggregation in ambient liquid water  
\cite{Chan,Ben,Lev,Oo,Jiang,Oka1,Bagchi}. 
\\  
(ii) When the  two  solvent components  
  interact with the solute very differently, 
we can have   $g_3^2 \chi\gg  1/n$ 
in Eq.(94)  (see Fig.3 in Ref.\cite{Oka1}).  
This tendency can appear even for $v_3^0 \sim 1/n$, 
but is amplified with increasing $v_3^0$. 
In this case, increasing $n_3$ results in 
 precipitation of solute-rich domains.  
This attractive interaction has been 
discussed for water-alcohol-hydrophobic solute 
(or  hydrotrope) \cite{Hori,Zemb1,Oka1}.\\ 
(iii) 
For solvent-polymer-colloid solutions, 
 Asakura and Oosawa calculated the effective 
potential between colloid particles 
due to   depletion of non-absorbing polymers  near  the 
colloid  surfaces  \cite{Oosawa,Vrij,Binder,Lek1}. 
In previous numerical calculations  
\cite{Ben1,Roij,Singh,Kino}, 
 the solute-solute radial distribution  function  
increased due to  the solvent-mediated 
attractions if the solute size 
exceeded  the solvent size.  
In these papers, the effective solute-solute 
potential range is on the order of the solute radius.

\vspace{2mm}
\noindent
{\bf 6.4. Solute-partial quantities   in terms of variances} \\
In this subsection,     we express the infinite-dilution limits of 
($\nu_{3i}, \zeta_3, g_3,   \nabla_T\nu_3)$ and 
  $({\bar v}_{3}^0, {\bar H}_{3}^0$)  in terms of the $T$-$\mu$ 
variances. First,  we note  
$I_{3i}/I_{33} \to n_i G_{3i}^0$ in Eq.(91), so  
\be 
\nu_{3i}= -  {\sum}_{j=1,2} n_jG_{j3}^0 
\lim_{n_3\to 0} I^{ji}~~(i=1,2),
\en 
from which $\zeta_3$ and $g_3$ in  Eqs.(80) and (81) are 
written as   
\bea 
&&\hspace{-10mm}
k_BT \kappa_T^m  \zeta_3= -{\sum}_{i=1,2}
 n_i v_i^m G_{3i}^0=  -\lim_{n_3\to 0}
 \frac{{\av{{\hat n}_3:{\hat \phi}}}}{n_3}, \\
&&\hspace{-10mm}   
 g_3= (G^0_{31}-G^0_{32})
\frac{n_1n_2}{n^2\chi}= -\lim_{n_3\to 0}
\frac{
{\av{{\hat n}_3:{\hat X}}}}{n_3\chi}. 
\ena 
Then, Eqs.(88) and (99) give the expression of 
${\bar v}_{3}^0$     originally derived by Ben Naim \cite{Naim3}: 
\be 
 {\bar v}_{3}^0 =
  k_BT\kappa_T^m -{\sum}_{i=1,2}
 n_i{\bar v}_i^m G_{3i}^0 . 
\en 

Next, from  Eqs.(32) and (86),   $M_3={\sum}_i n_i J_i I^{i3}$ 
tends to $k_BT(3/2-T\nabla_T \nu_{3})$
as $n_3\to 0$. Thus, using Eq.(91), we can relate 
$J_3^0={\lim}_{n_3\to 0}J_3$ and $\nabla_T\nu_3$ by  
\be 
J_3^0=  k_BT(3/2-T\nabla_T \nu_{3}) 
 - {\sum}_{i=1,2}\nu_{3i}n_i J_i^m .   
\en  
From Eqs.(36), (62),  and (89), 
 $\bar{H}_3^0$ is  written as  
\be 
{\bar H}_{3}^0  =J_3^0+ \lim_{n_3\to 0}\big[
 h {\bar v}_3  -  \av{{\hat e}:{\hat \phi}} +g_3\av{{\hat e}:{\hat X}}\big],   
\en 
where  $\av{{\hat e}:{\hat \phi}}\to J_1^m$ 
and $\av{{\hat e}:{\hat X}} \to 0$ as  $X\to 0$ in agreement with  Eq.(71).

\vspace{2mm}
\noindent
{\bf 7. Osmotic equilibrium in ternary fluids}\\
As in the biological situations 
 \cite{Pa,Smith,Smith1,Tim,Record,Shimizu}, 
we  suppose  a ternary solution composed of a mixture 
solvent ($i=1,2$) 
 and a dilute  solute ($i=3$), which is 
confined in a cell with a fixed volume $V$ 
 in  equilibrium with a large reservoir ($\gg V$). 
The two regions are separated by a semipermeable membrane. 
The  solvent  can pass through it 
to  acquire   common chemical potentials, 
$\mu_1$ and $\mu_2$,  in the two regions, 
while  there is no  solute  
  in the  reservoir. In the cell, 
   the solute density is increased  from 0 
to $n_3$   in osmotic equilibrium 
 at fixed $(T, \mu_1, \mu_2)$. 
Then,   all the deviations  are functions of $n_3$. 
In the biological papers,  the subscripts 2 and 3 
refer to  the solute and the cosolvent, 
respectively.  

The osmotic pressure     difference 
$\Pi= p(n_1, n_2, n_3,T)-p_0$ arises  across the membrane, 
where $p_0$ is the initial pressure. The reservoir pressure 
is unchanged from $p_0$. Using the Gibbs-Duhem relation 
and $d \mu_3=(k_BT/ I_{33})dn_3$, we obtain 
 the differential equation 
$d\Pi= n_3d\mu_3 = k_BT  (1+n_3 u_3^{\rm eff})dn_3$. 
Using $G_{33}^0$ in Eq.(95). we find  
the expansion,   
\be 
\Pi/k_BT= n_3 +n_3^2u_3^{\rm eff}/2 
= n_3- n_3^2 G_{33}^0/2,  
\en 
which is valid up to order $n_3^2$ \cite{Mc}. 
 For one-component solvents, Eqs.(94) and  (95)  
give $G_{33}^0=-U_{33}+ k_BT\kappa_T^m \zeta_3^2$ 
with Eq.(97) \cite{Koga,Ruck}.

We  also seek the 
 solute-induced changes in $n_1$ and $n_2$ in the cell. 
 As $n_3\to 0$, we have       
\be 
(\p n_i/\p n_3)_{T,\mu_1,\mu_2}=I_{3i}/I_{33} 
= n_i G_{3i}^0 ~~ (i=1,2).  
\en  
Therefore,  $n_1$ and $n_2$ change from their  initial values $n_i^0 $  by  
 \be
\Delta n_i=  n_i-n_i^0=
n_iG_{3i}^0 n_3 ~~(i=1,2).
\en 
Then, the solvent   
 volume fraction and the solvent  concentration change  by 
\bea 
&&\hspace{-12mm}
\Delta\phi= {\bar v}_1^m\Delta n_1+{\bar v}_2^m\Delta n_2 
=   -{\bar v}_3^0 n_3+ k_BT\kappa_T^m n_3,
. \\ ~~
&&\hspace{-12mm} \Delta X
= (G_{32}^0-G_{31}^0)(n_1n_2/n^2)n_3=  -\chi g_3 n_3.      
\ena  
where   Eq.(14)  is used  in Eq.(107) and  $g_3$ in Eq.(108) 
is given in Eq.(100).    
Note that Eq.(107)  is consistent with 
Eq.(11) since the pressure change  is equal 
to $\Pi\cong k_BTn_3$.  In Eq.(108),  
$|\Delta X|$ can much exceed $n_3/n$ 
for   $n|G_{32}^0-G_{31}^0|\gg 1$ 
   in the presence of 
 significant preferential 
solvationm,  which is the case for large hydrophobic solutes  
in water-alcohol solvents. 

In  biology \cite{Pa,Smith,Smith1,Tim,Record,Shimizu}, 
the degree of cosolvent  adsorption onto (or expulsion from) 
 solute particles  is represented by the 
preferential binding parameter (PBP) defined by  
\bea  
&&\hspace{-12mm}
\Gamma_{32}= \ppp{ m_2}{ m_3}{T, \mu_1, \mu_2}\hspace{-1mm} 
=\frac{ n^2k_BT}{n_1n_3} \ppp{ X}{ \mu_3}{T,\mu_1.\mu_2}
\\ 
&&\hspace{-6mm}=n_2 (G_{32}^0-G_{31}^0 )
=-({n^2}/{n_1})\chi g_3
=(n^2/{n_1})  {\Delta X}/{n_3},  
\ena  
where  $m_i= n_i/n_1$ (species molality) and  we use  
$dm_2/m_2= dn_2/n_2-dn_1/n_1=(n^2/n_1n_2) dX$ and   
 $dm_3\cong dn_3/n_1=n_3d\mu_3/k_BTn_1$. 
See  $g_3$ in Eqs.(81) and (100)  and $\Delta X$ in Eq.(108). 

 As another definition of PBP, the isobaric derivative   
 $\Gamma_{32}'=({\p m_2}/{\p m_3})_{T, p, \mu_2}$ 
has also been used \cite{Tim}, where $p$ is fixed instead of $\mu_1$. 
 As $n_3\to 0$, we find \cite{Record, Smith1}   
\be
\Gamma_{32}'
= \Gamma_{32} +{n_2}(1/{n_1}+ G_{11}^m-G_{12}^m) . 
\en   
Here,  $\Gamma_{32}'=w_2/w_3$ 
with 
 $w_i = (\p m_i /\p \mu_3)_{T.p,\mu_2}=
(\p m_i/\p \mu_3)_{T.\mu_1,\mu_2} - 
m_3(\p m_i/\p \mu_1)_{T.\mu_2,\mu_3}$.  
The second term in Eq.(111)  
  is independent of the solute properties and 
is  negligible  for large solutes with 
appreciable  preferential adsorption, where    
$n|G_{32}^0-G_{31}^0|\gg 1$ holds.

Finally, we calculate  the  solute-induced   change 
in the enthalpy $H$ from its initial value $H_0$ in the cell. 
At  fixed $(V,T,\mu_1,\mu_2)$, 
we have $dH=V(de+dp)$ with $dp= d\Pi$, 
so  Eqs.(20) and (105) yield the differential equation, 
\be
dH=  V{\sum}_i M_i (I_{i3}/ I_{33}) dn_3+ Vd\Pi
=V(J_3/k_BT+1)d\Pi. 
\en  
Thus,  in the  linear order in   $n_3 $, 
the osmotic enthalpy change is written in terms 
of $J_3^0$ as   
\be 
  \Delta H=H-H_0 =  V (J_3^0+k_BT)n_3.   
\en 
We also consider the enthalpy  change  $(\Delta H)_r$ in the reservoir. 
If the total solvent numbers in the two regions  
are fixed,  those  in the reservoir are changed  
by $-V\Delta n_i\cong -VI_{3i}$  from Eq.(106). 
Then,  from Eq.(34)  we find     
\be
 (\Delta H)_r= -V{\sum}_{i}{\bar H}_i^m I_{3i}
+ V{\bar H}_3^m I_{33}
= V({\bar H}_3^0 -k_BT^2\alpha_p^m- J_3^0)n_3. 
\en 
Therefore, the total enthalpy change is given by  
$\Delta H+ (\Delta H)_r=
[{\bar H}_3^0 +k_BT (1-T \alpha_p^m)]Vn_3$.

\vspace{-2mm}
\noindent
{\bf 8. Summary and remarks} \\
We have presented  a fluctuation  theory 
 in nonionic multi-component fluids. 
The integrals of the correlation functions  in the 
grand-canonical ensemble  have been   related  
to  the partial volumes ${\bar v}_i$, 
 the partial enthalpies ${\bar H}_i$, and 
other  thermodynamic derivatives. 

In Section 2, we have summarized the thermodynamic relations 
for   the partial quantities. 
The  differential equation (11) 
is  particularly useful, which reappears  in Eq.(47) 
for the thermal fluctuations. 
The  derivatives of the energy density $e$ with respect 
to the number densities $n_i$ are written as 
  $M_i$ at fixed $T$ and  $R_i$ at fixed $p$ 
in  Eqs.(18) and (19) and   are 
related to   ${\bar H}_i$ in  Eq.(22). 

In Section 3, we have defined the  variance 
 $\av{{\hat{\cal A}}:{\hat{\cal B}}}$ 
in the grand-canonical 
($T$-$\mu$) ensemble in Eq.(28).  
In addition to the Kirkwood-Buff integrals $G_{ij} $,  
we have  defined  the variances 
$n_iJ_i= \av{{\hat{e}}:{\hat{n}_i}}$ for 
 the energy density $\hat{e}({\bi r})$ 
and the number densities  $\hat{n}_i({\bi r})$. 
 Then, $J_i$   are expressed in  
terms of $M_i$,  $R_i$, and ${\bar H}_i$ 
 in Eqs.(32)-(35). We have  introduced 
 the fluctuations of the 
volume fraction   and the concentrations, 
$\delta{\hat \phi}({\bi r})$ and $\delta{\hat X}_i({\bi r})$,  
in Eqs.(38) and (39). The long-range correlations 
in the canonical ensemble has also been examined. 

In Section 4, we have introduced the fluctuations 
  in the temperature and the pressure,  
  $\delta{\hat T}({\bi r})$ and $\delta{\hat p}({\bi r})$, 
in Eqs.(45) and (46).    
which are  related to $\delta{\hat \phi}({\bi r})$ by Eq.(47) 
and   satisfy    variance relations in Eq.(48). 
We have clarified  
the relationship of  our theory to 
the  Greene-Callen  theory \cite{Callen1} 
and the  Landau-Lifshitz theory \cite{Landau-s}. 
We have  found general 
 variance relations of   the  stress tensor  
${\hat \Pi}_{\alpha\beta} ({\bi r})$ 
with  ${\hat n}_i({\bi r})$ and ${\hat e}({\bi r})$ 
in Eq.(51).   The long-range correlations 
in the $T$-$p$ ensemble has  been examined.

In Section 5, we have presented 
expressions for  ${\bar v}_i$ and ${\bar H}_i$ 
for binary mixtures in terms of 
 $J_i$ and $I_{ij}$  in   Eqs.(64) and (68). 
Their  critical behavior has also been discussed. 

In Section 6, we have treated solutions 
of a mixture solvent ($i=1, 2$) and a dilute solute ($i=3$), 
(i) The Helmholtz free energy density $f$  
has been  expanded with respect to the solute density 
$n_3$ in Eq.(74), where crucial is a solute-solvent 
interaction parameter $\nu_3(n_1, n_2, T)$. 
We have then expressed 
  ${\bar v}_3$ and ${\bar H}_3$ 
in terms of $\nu_3$ in  Eqs.(84) and (85). 
The critical behaviors of 
the infinite-dilution limits 
 ${\bar v}_3^0$ and ${\bar H}_3^0$ 
have been discussed.  
(ii) We have discussed the solvent-mediated 
solute-solute attraction   in Eqs.(91)-(94), which is 
 enhanced  for large solute sizes 
and/or significant asymmetry in 
the solute-solvent interactions. (iii) We have expressed  
${\bar v}_3$ and $\bar{ H}_3$ in terms of the 
$T$-$\mu$ variances.

In Section 7,  we have 
 calculated    the solvent density changes $\Delta n_i$ 
and  the enthalpy change $\Delta H$ 
due to addition of  a small amount of 
a  solute in the cell in  osmotic equilibrium with a reservoir. 
In particular, 
the solvent concentration change $\Delta X$ 
and  $\Delta H$ have been   expressed in terms of 
 $G_{3i}^0={\lim}_{n_3\to 0}G_{3i}$ 
and $J_3^0= {\lim}_{n_3\to 0}J_3$.

We make  remarks on future problems.\\ 
(i) We should construct a fluctuation theory  
for  electrolytes with   mixture solvents, 
where  ion dissociation 
and association occur particularly for weak acids and bases 
in water \cite{Ma}. 
Recently, we have examined the ion partial  volumes 
${\bar v}_i$ and the solvent-mediated ion-ion interactions 
in a waterlike solvent \cite{Oka5}. 
accounting for the 
hydration and the steric effect. 
We have also calculated the solvent-solvent  and 
solvent-ion correlations due to the ion solvation, 
which extend   over the inverse Debye wave number 
\cite{OnukiLong}.  \\
(ii) In chemical reactions, 
 heat release  (or absorption) 
and volume changes  
are  related to ${\bar H}_i$ and ${\bar v}_i$ 
 of the reactants and the products \cite{Myers,Callen,Landau-s}. 
In biology,  protein denaturation  has  been studied 
extensively   \cite{Pa,Smith,Smith1,Tim,Record,Shimizu}. 
 Thus, we should extend the present   theory 
 to include  the   reaction effects. 

\vspace{1mm}
 \noindent 
{\bf Data availability}: The data that support the findings of
this study are available within this  article.

\noindent{\bf Appendix A. Correlation functions in the 
grand-canonical ensemble   }\\
\setcounter{equation}{0}
\renewcommand{\theequation}{A\arabic{equation}}
For given $(T,\{ \mu \})$  at  a fixed volume, we write  the 
grand-canonical  distribution  as  \cite{Kubo} 
\be 
P_{\rm g} \propto 
\exp\Big[-\int d{\bi r}\Big({\hat e}({\bi r})B -{\sum}_i
{\hat n}_i({\bi r}) \xi_i\Big)
\Big], 
\en  
where $B=1/k_BT$ and $\xi_i=\mu_i/k_BT$. 
The average over $P_{\rm g}$ is written as $\av{\cdots}$. 
Then,  $e= \av{{\hat e}({\bi r})}$ and  $n_i = 
\av{{\hat n}_i({\bi r})}$ in the bulk. 
The logarithm of the grand-canonical partition function 
is $\omega V$ with  $\omega=p/k_BT$, 
so $\omega$  obeys  \cite{Gri,Kubo},  
\be 
d\omega= -edB+ {\sum}_i n_id\xi_i. 
\en 

If $B$ and $\xi_i$ are slightly changed by  
$\delta B$ and $\delta\xi_i$ homogeneously, $P_{\rm g}$ is changed by 
\be 
\delta P_{\rm g}
=\int d{\bi r}\Big[-\delta{\hat e}({\bi r})\delta B +
{\sum}_i \delta{\hat n}_i({\bi r}) \delta\xi_i\Big]  P_{\rm g}
\en 
Then, for any space-dependent fluctuating 
variable $\hat{\cal A}({\bi r})$,  
its average $\av{\hat{\cal A}}$ changes by  
\be 
\delta \av{\hat{\cal A}}= -\av{\hat{\cal A}:{\hat e}} \delta B 
+{\sum}_i \av{\hat{\cal A}:{\hat n}_i} \delta\xi_i, 
\en 	
using the variances in Eq.(28). 
As a function of $(B,\{ \xi\}$), 
the derivatives of $\av{\hat{\cal A}}$ are 
given by    
\be 
\frac{\p }{\p B}\av{\hat{\cal A}}= -\av{{\hat{\cal A}}:{\hat e}},~~
\frac{\p }{\p \xi_i}\av{\hat{\cal A}}
= \av{\hat{\cal A}:{\hat n}_i}. 
\en  
Treating $\omega$, 
$n_i$,  and $e$  as  functions of $(B, \{ \xi\})$, we have  
 \bea 
&&\hspace{-1cm}
 \av{{\hat n}_i:{\hat n}_j}=\p^2\omega/\p \xi_i\p\xi_j= 
\p n_i/\p \xi_j=I_{ij},\\
&&\hspace{-1cm}
\av{{{\hat e}}:{\hat n}_i}=-\p^2\omega/\p \xi_i\p B
=-{\p n_i}/{\p B}= {\p  e}/{\p \xi_i}, \\
&&\hspace{-1cm} \av{{{\hat e}}:{\hat e}}=\p^2\omega/\p B^2
=-{\p e}/{\p B},  
\ena 
where Eq.(A6) leads to Eq.(30). On the other hand, 
${\p  e}/{\p \xi_i}$ in  Eq.(A7) 
is equal to ${\sum}_j M_jI_{ji}$ from Eq.(18), 
leading to   Eq.(32). Next,  we rewrite Eq.(A8) as    
\be 
\av{{{\hat e}}:{\hat e}}=-  
\ppp{ e}{ B}{\{ n\}}- {\sum}_i M_i\ppp{n_i}{B}{\{\xi\}},
\en 
where the first term is $k_BT^2C_V$ 
and $({\p n_i}/{\p B})_{\{\xi\}}$ in the second term is 
$-\av{{{\hat e}}:{\hat n}_i}= -{\sum}_{j}  M_j I_{ij}$ 
from Eq.(A7). Thus, we find     Eq.(35).

The  local stress tensor  ${\hat \Pi}_{\alpha\beta}({\bi r})$ 
is microscopically defined with  $\av{{\hat \Pi}_{\alpha\beta}}=
\delta_{\alpha\beta}p$ 
(see the sentences above Eq.(51)), 
so Eq.(A5) gives  
\be
\delta_{\alpha\beta}\frac{\p p}{\p B}
=- \av{{\hat \Pi}_{\alpha\beta}:{\hat e}} ,~~
 \delta_{\alpha\beta}\frac{\p p}{\p \xi_i} =
\av{{\hat \Pi}_{\alpha\beta}:{\hat n}_i}  ,
\en 
which lead to Eq.(51)  
from   Eq.(A2). 
Generally,  for any microscopically defined 
$\hat{\cal A}({\bi r})$, 
we can determine   a  
variable $\hat{\cal A}_{\alpha\beta}({\bi r})$ with  
 $\av{{\hat \Pi}_{\alpha\beta}:\hat{\cal A}}
= k_BT \av{\hat{\cal A}_{\alpha\beta}}$    
(see Appendix 1A in Ref.\cite{Onukibook}).

Furthermore, we can   rewrite $\delta P_{\rm g}$ in 
Eq.(A3)  as \cite{Onukibook} 
\be
\hspace{-0.4mm} 
\delta P_{\rm g}= B\int\hspace{-1mm}
 d{\bi r}\Big[{n\delta{\hat \sigma}{\delta T} }
 +  \frac{\delta{\hat n}}{n} \delta p  
 + {\sum}_{i} n\delta{\hat X}_i   
{\delta\mu_i}\Big]  {P_{\rm g}}.
\en 
where  $\delta{\hat X}_i$ are the fluctuating 
 concentrations  in Eq.(39), and  
$\delta p= s\delta T 
+ {\sum}_i n_i\delta \mu_i$ is the pressure deviation. 
We define 	
   the fluctuating entropy deviation per particle  by    
\bea  
&&\hspace{-11mm}
\delta{\hat \sigma}({\bi r})=[\delta{\hat e}- {\sum}_i
	\mu_i\delta{\hat n}_i]/nT - s {\sum}_i\delta{\hat n}_i/n^2 
\nonumber\\
&&= {\sum}_i {\bar S}_i \delta{\hat X}_i+ 
C_p \delta{\hat T}/nT- \alpha_p \delta{\hat p}/n  . 
\ena 
In the second line,  $\delta{\hat T}$ and $\delta{\hat p}$ 
are the temperature and 
pressure fluctuations   in  Eqs.(45) and (46), so  
 Eq.(48) gives  
$\av{{\hat T}:{\hat \sigma}}= k_BT/n$ and $ 
\av{{\hat p}:{\hat \sigma}}=0$. 
From  Eq.(A11) we  
express the average deviation $\delta \av{\hat{\cal A}}$
in terms of    the thermodynamic deviations 
$\delta T$, $\delta p$, and $\delta\mu_i$ as  
\be 
k_BT 
\delta \av{\hat{\cal A}}= {
\av{\hat{\cal A}:{\hat \sigma}}}n\delta T   
+{\av{\hat{\cal A}:{\hat n}} }{\delta p}/{n}
+{\sum}_{i} {\av{\hat{\cal A}:{\hat X}_i}} 
{n\delta\mu_i},  
\en 	
which is equivalent to Eq.(A4) in the linear order.

 For binary mixtures, we set   
$\delta {\hat X}= \delta {\hat X}_2$ 
and $\delta \Delta = \delta\mu_2-\delta \mu_1$. 
Then, the last term in Eq.(A13) becomes 
$ \av{\hat{\cal A}:{\hat X}}n\delta\Delta$.  
At fixed $(T,p)$,  we thus  find       
\be 
k_BT 
\ppp{\av{\hat{\cal A}}}{\Delta}{T,p} =
\ppp{\av{\hat{\cal A}}}{ X}{T,p}{n\chi}   
={\av{\hat{\cal A}:{\hat X}}}{n},   
\en 
leading to     Eqs.(63) and (70) 
for   $\hat{\cal A}={\hat X}$ 
and  ${\hat e}$. 
Since  $(\p w/\p T)_{p,X}=(\p w/\p T)_{p,\Delta}- 
(\p w/\p \Delta)_{T,p}\av{\hat{\sigma}:\hat{X}}/\chi$ 
for any $w$, 
 $\alpha_p$ and $C_p$   in Eq.(9)  are given by 
  \cite{Onukibook}
\bea 
&& \hspace{-15mm} 
k_BT \alpha_p= -\av{\hat{\sigma}:\hat{n}}
+ \av{\hat{\sigma}:\hat{X}}\av{\hat{n}:\hat{X}}
/\chi,\\  
&& \hspace{-15mm} 
k_B C_p/n^2=\av{\hat{\sigma}:\hat{\sigma}}
- \av{\hat{\sigma}:\hat{X}}^2/\chi.   
\ena  
Here,  Eqs.(A15) and (A16) 
give $\av{\Delta S\Delta V}/V$  
and $\av{(\Delta S)^2}/V$, respectively, 
 in the  $T$-$p$ ensemble  from Eq.(60), where 
 $\Delta S= n\int_V d{\bi r}\delta{\hat\sigma}({\bi r})$.  
is the total  entropy change. We can also derive 
 $\kappa_T$ in Eq.(67) from Eq.(A13).

\noindent{\bf Appendix B.  Expansions 
 up to order $n_3$ }\\
\setcounter{equation}{0}
\renewcommand{\theequation}{B\arabic{equation}}
In this appendix, we give  some additional  
 results   for a ternary fluid containing  a dilute solute with 
density $n_3$.
From Eqs.(15), (26), and (74) 
$\kappa_T$,  $ \beta_V$, and $\alpha_p=\kappa_T \beta_V$ 
 are expanded from their solvent values in Eq.(87) 
 up to order $n_3$ in terms of $\nu_3$ as 
\bea 
&&\hspace{-1cm}
\kappa_T= \kappa_T^m + \Big[2\zeta_3+1+    
{\sum}_{i,j=1,2} \nu_{3ij}n_in_j\Big]k_BT\kappa_T^2 n_3,\nonumber\\
&& \hspace{-1cm} 
\beta_V=\beta_V^m +(1+ \zeta_3+ T\nabla_T\nu_{3})k_B n_3, 
~~\alpha_p/ \alpha_p^m = \kappa_T/\kappa_T^m+ 
\beta_V/\beta_V^m-1,
\ena 
where $\nu_{3ij} =(\p^2 \nu_3/\p n_i \p n_j)_T$ $(i,j=1,2)$.

Next, we present  the solvent ${\bar v}_i$ and 
${\bar H}_i$ ($i=1,2$) up to order $n_3$ 
from Eqs.(14), (22),  (84), (85), and (86) as  
\bea 
&&\hspace{-10mm}
{\bar v}_i={\bar v}_i^m (1- {\bar v}_3n_3) +(\delta_{i2}-X)
k_BT\kappa_T\frac{n_3}{n}\ppp{\zeta_3}{X}{T,p},\\
&&\hspace{-10mm}
{\bar H}_i= {\bar H}_i^m - k_BT^2(\nabla_T\nu_3) n_3+ 
T(\beta_V {\bar v}_i-\beta_V^m {\bar v}_i^m) .
\ena 
These relations assure  ${\sum}_i n_i {\bar v}_i=1$ 
and ${\sum}_i n_i {\bar H}_i=h$ up to order $n_3$. In the latter  
 we use   
\be
h=\lim_{n_3\to 0}h + {\bar H}_3n_3+ k_BTn_3(1+ \zeta_3 )
(1-T\alpha_p^m),  
\en  
which follows from Eqs.(77) and (85),

\end{document}